\documentclass[twocolumn,tighten,times]{aastex63}
\usepackage{graphicx}
\usepackage{amssymb}
\usepackage{amsbsy}
\usepackage{amsmath}
\usepackage{amsfonts}
\usepackage{mathrsfs}
\usepackage{color}
\usepackage{multirow}
\usepackage{bm}

\DeclareMathAlphabet{\mathbi}{OT1}{ptm}{bx}{it}
\SetMathAlphabet\mathbi{bold}{OT1}{ptm}{bx}{it}

\defcitealias{Gravity2018}{G18} 	  	
\defcitealias{Wang2020}{W20} 

\received{2021 August 7}
\revised{2021 October 27}
\accepted{2022 January 10}
\submitjournal{\it The Astrophysical Journal}

\shorttitle{SARM: the SMBH Mass and Distance in 3C 273}
\shortauthors{Li et al.}
 
\begin{document}

\title{\bf\large Spectroastrometry and Reverberation Mapping: the Mass and Geometric Distance 
of the Supermassive Black Hole in the Quasar 3C 273}

\author[0000-0001-5841-9179]{Yan-Rong Li}
\affiliation{Key Laboratory for Particle Astrophysics, Institute of High 
Energy Physics, Chinese Academy of Sciences, 19B Yuquan Road, 
Beijing 100049, China}
\author[0000-0001-9449-9268]{Jian-Min Wang}
\affiliation{Key Laboratory for Particle Astrophysics, Institute of High 
Energy Physics, Chinese Academy of Sciences, 19B Yuquan Road, 
Beijing 100049, China}
\affiliation{National Astronomical Observatories of China, Chinese 
Academy of Sciences, 20A Datun Road, Beijing 100012, China}
\affiliation{School of Astronomy and Space Science, University of Chinese Academy of Sciences, 
19A Yuquan Road, Beijing 100049, China}
\author[0000-0003-4042-7191]{Yu-Yang Songsheng}
\affiliation{Key Laboratory for Particle Astrophysics, Institute of High 
Energy Physics, Chinese Academy of Sciences, 19B Yuquan Road, 
Beijing 100049, China}
\author{Zhi-Xiang Zhang}
\affiliation{Department of Astronomy, Xiamen University, Xiamen, Fujian 361005, China}
\author[0000-0002-5830-3544]{Pu Du}
\author{Chen Hu}
\author{Ming Xiao}
\affiliation{Key Laboratory for Particle Astrophysics, Institute of High 
Energy Physics, Chinese Academy of Sciences, 19B Yuquan Road, 
Beijing 100049, China}

\email{liyanrong@mail.ihep.ac.cn, wangjm@mail.ihep.ac.cn}

\begin{abstract}
The quasar 3C~273 has been observed with infrared spectroastrometry (SA) on broad Pa$\alpha$ 
line and optical reverberation mapping (RM) on broad H$\beta$ line. SA delivers information about
the angular size and structure of the Pa$\alpha$ broad-line region (BLR), while 
RM delivers information about the physical size and structure of the H$\beta$ BLR. Based on
the fact that the two BLRs share the mass of the 
supermassive black hole (SMBH) and viewing inclination, a combination of SA and velocity-resolved RM 
(SARM) thereby allows us to simultaneously determine the SMBH mass and geometric distance 
through dynamically modeling the two BLRs. We construct a suite of dynamical models with 
different geometric configurations and apply a Bayesian approach to obtain the parameter inferences. 
Overall the obtained masses and distances are insensitive to specific BLR configurations 
but more or less depend on parameterizations of the vertical distributions. 
The most probable model, chosen in light of the Bayes factor, yields an angular-size distance of 
$\log\,(D_{\rm A}/{\rm Mpc}) = 2.83_{-0.28}^{+0.32}$ and SMBH mass of 
$\log\,(M_\bullet/M_\odot)=9.06_{-0.27}^{+0.21}$, which agrees with the relationships between SMBH masses and 
bulge properties. The BLRs have an inclination of $5_{-1}^{+1}$ degrees, consistent with that of the 
large-scale jet in 3C~273. 
Our approach reinforces the capability of SARM analysis to measure 
SMBH mass and distance of AGNs even though SA and RM observations are undertaken with different emission lines and/or in 
different periods.
\end{abstract}

\section{Introduction}
Active galactic nuclei (AGNs) and quasars are those the brightest objects at cosmic distances that 
are believed to be powered by matter feeding into supermassive black holes (SMBHs) residing at their centers. 
One of hallmark properties of their electromagnetic radiations is the presence of prominent broad emission lines 
with typical widths of several thousands kilometers per second (e.g., \citealt{Netzer2013}).  It is long known 
that the regions responsible for broad emission lines, the so called broad-line regions (BLRs), must be compact 
and located well within the centers of AGNs and quasars as as to produce sufficiently large Doppler shifting velocities 
and explain the observed line widths (e.g., \citealt{Woltjer1959, Burbidge1963, Rees1977}). These broad emission lines 
have been of central importance to unveil gaseous environments and dynamics in the vicinity of SMBHs.
However, because of limited spatial resolution of observations, it had always been challenging to determine fine 
structures and dynamics of BLRs in AGNs and quasars.

The reverberation mapping (RM) technique, originating from the pioneering works of \cite{Bahcall1972} and 
\cite{Blandford1982}, swaps spatial resolution for time resolution by measuring temporal responses of variations of 
broad emission lines to those of the driving continuum emitted by the accretion disk (\citealt{Peterson1993}).
These responses show time delays due to the light-travel time from the accretion disk to the BLR and thus encode
information about BLR structure and kinematics perpendicular to the iso-delay surfaces. 
Starting in the late 1980s, RM observation campaigns
have significantly advanced our understanding of BLRs and most importantly, have
led to the establishment of RM as a promising method for measuring SMBH masses in AGNs 
(e.g., \citealt{Kaspi2000, Bentz2013, Du2019}; see also a review in \citealt{Peterson2014}).

On the other hand, after decades of persistent endeavors, the GRAVITY instrument onboard the Very Large Telescope Interferometer 
(VLTI; \citealt{Gravity2017}) successfully resolved, for the first time, the Pa$\alpha$ BLR in the quasar 3C~273 
with an angular resolution down to $\sim$10 micro arcseconds  (\citealt[hereafter \citetalias{Gravity2018}]{Gravity2018}). 
This achievement marked a giant step towards the era of high angular resolution astronomy in the field of AGNs 
and most importantly, ushered in the practical feasibility of applying the spectoastrometry (SA) technique to 
understand BLR physics (\citealt{Beckers1982, Bailey1998}). The GRAVITY/VLTI instrument measures SA signals of the BLR, 
namely,  astrometry along with wavelength (or velocity), through interferometric phases. 
Because different parts of the BLR have different Doppler shifting velocities and photocenters (or angular displacements),
SA can therefore resolve angular structures of the BLR perpendicular to the line of sight (LOS; 
e.g., \citealt{Rakshit2015, Gravity2020a}). Besides through interferometry, 
SA can also be applied through spectroscopy with the forthcoming next-generation single-aperture telescopes, 
such as the Extremely Large Telescope, the Thirty Meter Telescope, and the Giant Magellan Telescope, 
all which are expected to attain an angular resolution of several tens of micro arcseconds (e.g., \citealt{Stern2015, Bosco2021}).

With the successful SA observation of \citetalias{Gravity2018}, \citet[hereafter \citetalias{Wang2020}]{Wang2020}
made the first effort to integrate the SA technique (in spatial domain) with the reverberation mapping (RM) technique 
(in time domain) and conducted the joint SA and RM (hereafter SARM)
analysis on the data of 3C 273 (\citetalias{Gravity2018}; \citealt{Zhang2019}).
A second case of SARM analysis was recently conducted by \cite{Gravity2021} upon 
the nearby galaxy NGC 3783.
The capability of such SARM analysis is twofold. First, SA and RM are complementary in the sense that they
probe different dimensions of BLR structures. 
SARM analysis thus leverages this property and offers new physical information about BLRs 
(compared to sole SA or RM analysis). Second, the linear BLR structures from RM 
together with angular BLR structures from SA naturally constitute a basic probe to geometric distances of AGNs and quasars. 
In this regard, the SARM approach paves an elegant pathway for the long-term epic pursuits of using AGNs and quasars 
for cosmology, which date back to 1960s soon after the discovery of the first quasar 3C 273 (e.g., 
\citealt{Sandage1965, Hoyle1966, Longair1967, Schmidt1968, Bahcall1973,
Baldwin1977, Elvis2002, Watson2011, Wang2013, Honig2014, LaFranca2014, Risaliti2019, Gravity2021b}).

For 3C~273, SA observed the infrared Pa$\alpha$ line (\citetalias{Gravity2018}) whereas RM observed the optical 
H$\beta$ line (\citealt{Zhang2019}). It is known that profiles of H$\beta$ and Pa$\alpha$ lines in AGNs are generally 
not consistent in light of both widths and shapes (\citealt{Landt2008, Kim2010, Durre2021}). Specifically, 
\cite{Kim2010} reported that the full widths at half maximum (FWHMs) of H$\beta$ lines are systematically broader 
than those of Pa$\alpha$ lines by about $\sim30\%$. This clearly implies that the respective 
BLRs of the two lines are different in some aspects (e.g., optical depths). 
Because only using 1D RM\footnote{Here, 
1D RM means that only velocity-integrated fluxes of the broad emission line
are used in RM analysis and the velocity dimension is not taken into account. Oppositely, 2D (or velocity-resolved)
RM means that the additional velocity dimension is included.} data, the SARM analysis in \citetalias{Wang2020} needed to
presume a common BLR for the two lines and neglected their possible differences. 
It is therefore highly worth extending the work of \citetalias{Wang2020} by developing a more generic framework. 
As we show below, through combining SA data and 2D (or velocity-resolved) RM data in a way that the corresponding BLRs only share 
some physical parameters (such as black hole mass and inclination), 2D SARM analysis can resolve not only the problem 
of different emission lines in the two data sets, but also the issue that SA measures emissivity-weighted photocenters 
of the BLR, whereas RM measures responsivity-weighted parts of the BLR (e.g., \citealt{Goad1993}). Besides, such a new 
approach of SARM analysis does not even require SA and RM data to be observed in the same periods, therefore greatly 
strengthening the applicability of SARM analysis.

The paper is organized as follows. 
Section~\ref{sec_data} compiles SA and RM data as well as compares the profiles of the extracted H$\beta$ 
and Pa$\alpha$ lines. Section~\ref{sec_blr} introduces BLR dynamical modeling on SA and RM data and constructs 
a suite of BLR models. In Section~\ref{sec_bayes}, we detail the procedure for joint SA and 2D RM analysis.
In Section~\ref{sec_results}, we summarize the main results of our joint analysis. 
Discussions and conclusions are given in Sections~\ref{sec_discussion} and \ref{sec_conclusion}, respectively. 

Throughout the paper, we use the redshift $z=0.15834$ of 3C 273 retrieved from the NASA/Infrared Process and 
Analysis center (IPAC) Extragalactic Database, with which we convert the wavelength to the rest frame and 
correct for the cosmological time dilation.
The Galactic extinction along the direction of
3C~273 is corrected using a color excess of $E(B-V)=0.018$ and the \cite{Fitzpatrick1999} parameterization.

\section{Data}\label{sec_data}
\subsection{Spectroastrometric Data}
The SA data of 3C~273 were reported by \citetalias{Gravity2018}, which made observations using the GRAVITY/VLTI
(\citealt{Gravity2017}) on eight nights between July 2017 and May 2018.
The instrument measures the interferometric visibility amplitudes and phases on each of six baselines (telescope pairs)
by coherently combining the light from the four 8m telescopes. \cite{Gravity2020} reported the measured 
visibility amplitudes for 3C~273, which are dominated by the visibilities of the continuum emission from 
hot dusts. \citetalias{Gravity2018} averaged the exposures of adjacent epochs to reduce measurement noises and published 
the differential phase curves (relative to a reference continuum wavelength) as a function of wavelength 
for six baselines on four (averaged) epochs (see extended data Figure 1 therein). 
To calculate differential phases with a BLR dynamical model, the Pa$\alpha$ profile normalized by the underlying 
continuum flux is also needed. We use the mean Pa$\alpha$ profile over the eight epochs between 
July 2017 and May 2018 published by \citetalias{Gravity2018}. We use a Gaussian to model the GRAVITY spectral broadening 
with a standard deviation of 235~km~s$^{-1}$ (\citetalias{Gravity2018}).

\begin{figure}[t!]
\centering 
\includegraphics[width=0.48\textwidth]{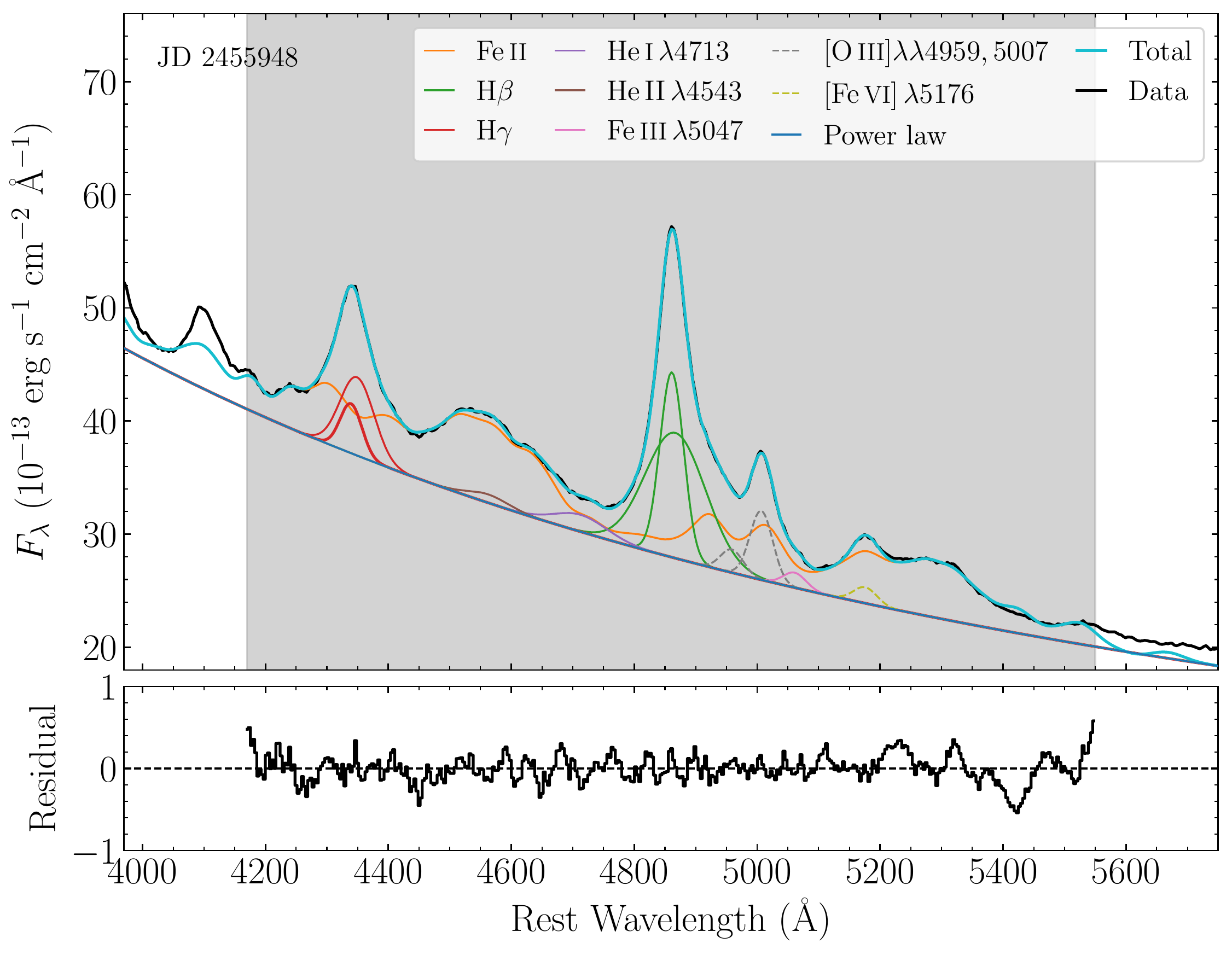}
\caption{An example for a multi-component decomposition of the spectrum at JD 2455948. Grey shaded area represents the wavelength 
window for the spectral decomposition. }
\label{fig_spec}
\end{figure}

\subsection{Reverberation Mapping Data}
\label{sec_rm}
The RM campaign of 3C 273 was reported by \cite{Zhang2019}, which synthesized the spectroscopic
and photometric data from the Steward Observatory spectropolarimetric monitoring 
project\footnote{The website is at \url{http://james.as.arizona.edu/~psmith/Fermi/}} (\citealt{Smith2009}) and the super-Eddington 
accreting massive black hole project (\citealt{Du2014}).
These two projects employed different kinds of spectrographs and the yielded data differed in  
spectral resolutions and flux calibrations. The majority of the spectroscopic data come from the Steward Observatory, therefore,
for the sake of data homogeneity, we only use the Steward Observatory's data.
There were in total 386 spectra during the period between November 2008 and July 2018, out of which 
301 spectra were calibrated to match the $V$-band magnitude taken on the same night and the rest 75 spectra 
could not be calibrated as such because of lacking contemporaneous photometry. 
We further discard 23 spectra taken under an airmass $\geqslant2.0$.
In addition, there were double exposures in eight epochs of the remaining spectra. 
We combine those double exposures on the same night into one and 
finally get 270 spectra. 

To extract the broad H$\beta$ emission line from the spectra, we implement a spectral decomposition scheme 
described in \cite{Hu2012} and \cite{Barth2013}, using the {\texttt{DASpec}} software\footnote{Available at 
\url{https://github.com/PuDu-Astro/DASpec}.}, which determines all spectral components simultaneously 
by minimizing the $\chi^2$ via the Levenberg–Marquardt technique. 
The spectral fitting window is set to be 4170-5550~{\AA} (in rest frame).
We include the following spectral components: 1) a single power law for the featureless continuum; 
2) a \ion{Fe}{2} template from \cite{Boroson1992} to represent the optical blended \ion{Fe}{2} emissions; 
3) single Gaussians for the [\ion{O}{3}]$\lambda\lambda$4959, 5007. 
The fluxes of [\ion{O}{3}]$\lambda$4959 is
fixed to be one-third of those of [\ion{O}{3}]$\lambda$5007 (e.g., \citealt{Peterson2004}); 
4) double Gaussians for the broad 
H$\beta$ and H$\gamma$ lines; 5) a single Gaussian for the broad \ion{He}{1} $\lambda$4713 line (e.g., \citealt{Veron2002}).
This line is strongly blended with the blue wing of the H$\beta$. To reduce the degeneracy, 
we fix a zero velocity shift for this line component. The narrow-line components for the H$\beta$ and 
H$\gamma$ lines are not included because the [\ion{O}{3}] lines are quite weak, indicating that the narrow H$\beta$ 
component is negligible.

\begin{figure*}[th!]
\centering 
\includegraphics[width=0.45\textwidth]{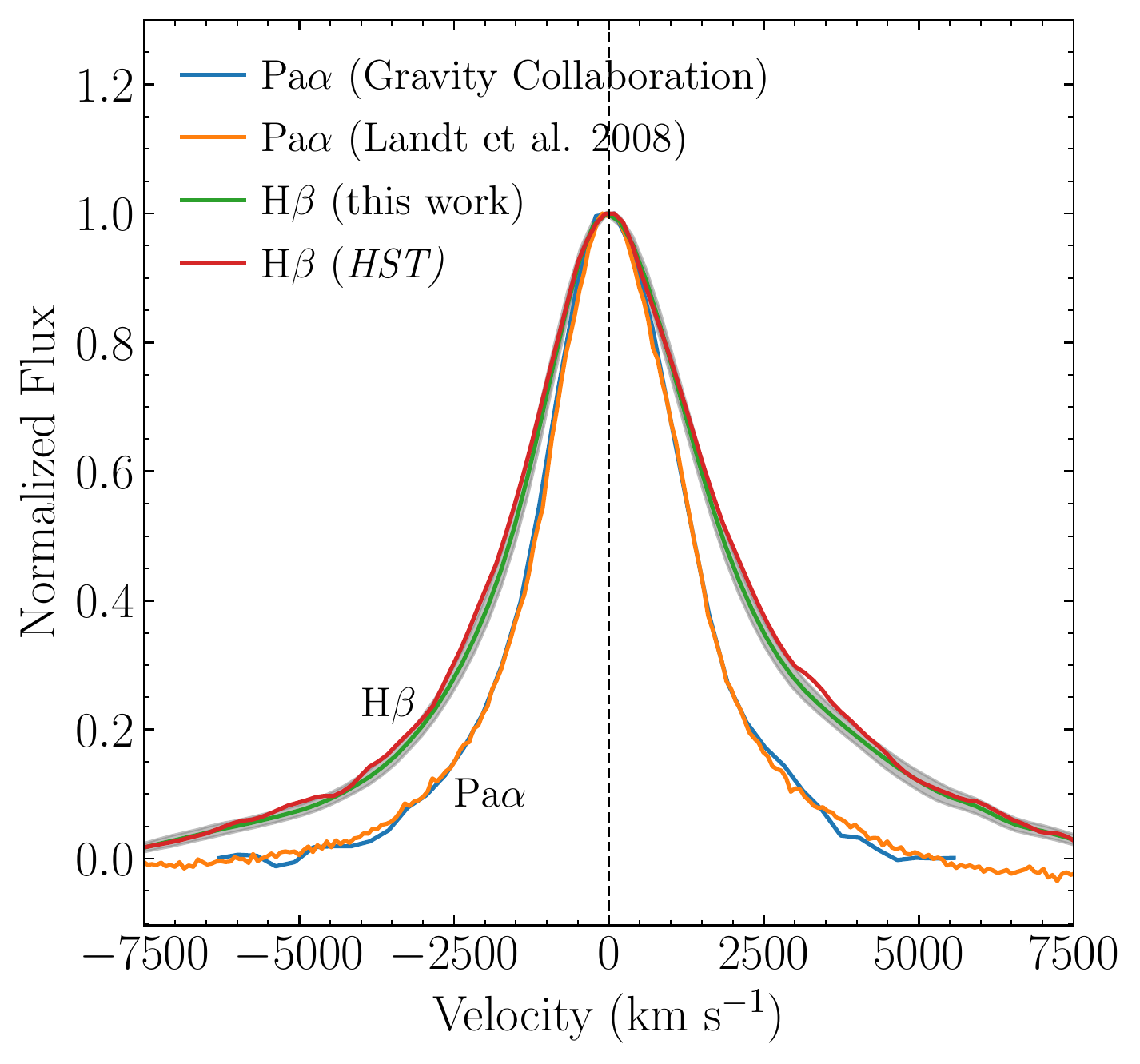}~~~
\includegraphics[width=0.45\textwidth]{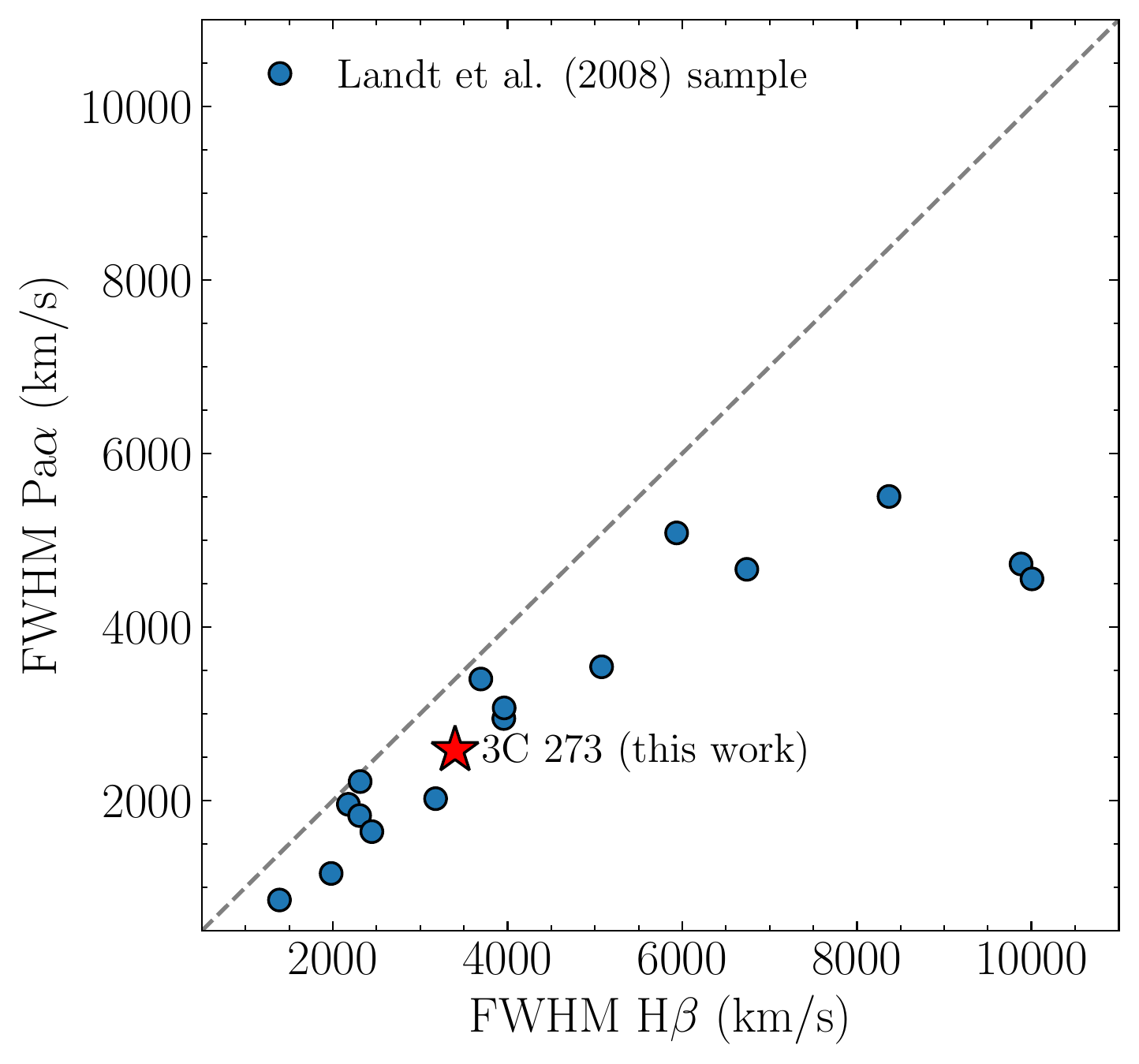}
\caption{(Left) A comparison between the broad H$\beta$ and Pa$\alpha$ profiles of 3C 273. Grey shaded area represents 
the standard deviation of the mean H$\beta$ profile from the RM observations (see the text for a detail). 
(Right) Pa$\alpha$ FWHMs with H$\beta$ FWHMs for a sample of AGNs compiled
by \cite{Landt2008}. The location of 3C~273 from this work is highlighted by an asterisk. Note that \cite{Landt2008} used a 
different scheme for extracting the broad line components, but there is a general consensus that H$\beta$ lines are systematically 
broader than Pa$\alpha$ lines.}
\label{fig_prof}
\end{figure*}

Besides the above components, there are additional significant components around 4543~{\AA}, 5047~{\AA}, 
and 5176~{\AA}, which might either correspond to \ion{He}{2}, \ion{Fe}{3}, and [\ion{Fe}{6}] emissions, respectively, 
or be ascribed to the \ion{Fe}{2} emissions that are not accounted in the template.
We use single Gaussians to represent these components. In particular, the profile width and velocity offset 
of the [\ion{Fe}{6}] line are tied to those of the [\ion{O}{3}]$\lambda$5007.
The host starlight component is not included as its contribution to the total flux 
is estimated to be less than 6\% from the host image decomposition (see \citealt{Zhang2019} for a detail).
We note that compared to the spectral decomposition scheme in \cite{Zhang2019}, we additionally include 
broad H$\gamma$, \ion{He}{1}~$\lambda4713$ line, and Guassian components around 4543~{\AA} and 5047~{\AA}
to further improve the spectral fitting.

After running the above spectral decomposition scheme over all spectra, we obtain the wavelength offset
of the [\ion{O}{3}]$\lambda$5007 for each spectrum. We find that the mean offset is almost equal to 
zero, but the dispersion is as large as 2.6~{\AA}.
It is therefore necessary to correct for the wavelength offset by aligning the center wavelengths of the [\ion{O}{3}]$\lambda$5007. 
We finally rerun the above spectral decomposition scheme again 
and obtain the broad H$\beta$ profile by subtracting all the other components from each spectrum.
In Figure~\ref{fig_spec}, we show an example for the decomposition of the spectrum at JD 2455948.

We estimate the spectral broadening arising from the instrument and seeing by comparing 
the [\ion{O}{3}]$\lambda5007$ line width to the intrinsic width that can be deduced from 
otherwise observations with well-determined echelle line spread functions.
Following \cite{Zhang2019}, we use the spectrum\footnote{The spectrum is extracted from the dataset \texttt{O44301020} of the {\it HST} archives.} 
of 3C~273 observed on January 31, 1999 using grating G750L of the Space Telescope Imaging Spectrograph 
(STIS) onboard the {\it Hubble Space Telescope (HST)} and apply the above spectral decomposition scheme to 
determine the observed FWHM of the [\ion{O}{3}]$\lambda5007$ line $(1.97\pm0.03)\times10^3~{\rm km~s^{-1}}$.
By subtracting the STIS' instrumental resolution of 439~$\rm km~s^{-1}$ in quadrature, we derive the 
intrinsic [\ion{O}{3}]$\lambda5007$ line width to be $(1.92\pm0.03)\times10^3~{\rm km~s^{-1}}$.
We then subtract this intrinsic width from the observed [\ion{O}{3}]$\lambda5007$ line widths of our RM spectra 
to obtain the spectral broadening at each epoch, which generally ranges from $\sim$800~$\rm km~s^{-1}$ to $\sim$1400~$\rm km~s^{-1}$
with a mean of 1121~$\rm km~s^{-1}$ and standard deviation of 203~$\rm km~s^{-1}$.

\subsection{A Comparison of the Broad H$\beta$ and Pa$\alpha$ Lines}
In the left panel of Figure~\ref{fig_prof}, we compare the profiles of the broad H$\beta$ from the mean 
spectrum of the RM data and Pa$\alpha$ lines from the SA data.
We also superimpose the broad H$\beta$ and Pa$\alpha$ lines extracted from the {\it HST} spectrum observed on January 31, 1999 (see 
Section~\ref{sec_rm})
and the infrared spectrum of 3C~273 observed by \cite{Landt2008} on June 12, 2006, respectively.
Both the normalized H$\beta$ and Pa$\alpha$ line profiles do not show noticeable changes between epochs spanning longer than a decade,
plausibly implying that the BLRs in 3C~273 are dynamically stable over such a timescale.

The mean H$\beta$ profile has an FWHM of $\sim3200~{\rm km~s^{-1}}$ while the Pa$\alpha$ profile has an 
FWHM of $\sim2500~{\rm km~s^{-1}}$, narrower than the H$\beta$ profile by a factor of $\sim30\%$. 
In the right panel of Figure~\ref{fig_prof}, we plot Pa$\alpha$ and H$\beta$ line widths for an AGN sample
measured by \cite{Landt2008} and superimpose the widths of 3C~273 from this work for sake of comparison. 
We note that \cite{Landt2008} used a different scheme for extracting the broad line components.
Nevertheless, we can find a general consensus that H$\beta$ lines are systematically broader than Pa$\alpha$ lines
(see also \citealt{Kim2010}).
In addition, the left panel of Figure~\ref{fig_prof} illustrates that the H$\beta$ profile 
is slightly asymmetric, with a minor excess in the red wing. By contrast, the P$\alpha$ profile is relatively symmetric.

The above factors indicate that the geometry and/or kinematics of the 
H$\beta$ and P$\alpha$ emission regions might not be the same.
Although we note that these two lines share an upper Hydrogen atomic level,  complicated dependences of the respective combination 
coefficients on density and temperature distributions of the BLR gas might be a contributing factor to the 
different emission regions (e.g., \citealt{Osterbrock2006}, Chap. 4).
In subsequent analysis, we therefore use two independent sets of parameters to describe the H$\beta$
and Pa$\alpha$ BLRs, except that they share the inclination and central black hole mass.

\section{Broad-line Region Modeling}\label{sec_blr}
\subsection{Basic Assumptions}\label{sec_assumptions}
Before constructing BLR models, we elaborate on necessary {\it ansatzs} for the sake of simplifying calculations: 
\begin{itemize} 
\item BLRs are composed of 
a large number of discrete, point-like clouds, which move in the gravitational well of
the central black hole and reprocess the central ionizing emissions to line emissions with light-travel 
delays. Any shadowing among clouds is neglected (\citealt{Pancoast2014}).

\item BLRs are stable in both structure and kinematics during the period under consideration,
that is to say, all variability in emission lines are a result of reprocessing continuum variability so that 
RM is right applicable. The BLR ``breathing'' effects are not included (e.g., \citealt{Cackett2006}) 
considering the mild fractional variability $\lesssim10\%$ of the 5100~{\AA} continuum and H$\beta$ line for 3C~273
(\citealt{Zhang2019}). This is also supported by the almost unchanged H$\beta$ and Pa$\alpha$ profiles over more than 
a decade as illustrated in Figure~\ref{fig_prof}.

\item The central ionizing source is compact and has a point-like geometry so that its emissions are isotropic.
The ionizing continuum is not available and the optical continuum is used as a substitute by 
assuming that the two bands of continuum emissions are linearly correlated with a negligible inter-delay (compared to 
the time delays of BLRs). The assumption of a linear correlation would be reasonable provided 
that the spectral energy distribution of the ionizing 
source received by BLR clouds does not change significantly.

\item 
Detailed gas physics (such as photoionization) are not included and each cloud's emissions are simply proportional 
to the incident continuum fluxes with a nonlinear parameterization (see also \citealt{Li2013}).
Specifically, this means that changes in line emissivity $\epsilon(t)$ is related to changes in incident continuum flux $f_{c}(t)$
for clouds at a radius $r$ with a time delay $\tau$
\begin{eqnarray}
\frac{\dot\epsilon(r, t)}{\epsilon(r, t)} = (1+\delta)\frac{\dot f_c(r, t)}{f_c(r, t)} = (1+\delta)\frac{\dot F_c(t-\tau)}{F_c(t-\tau)},
\end{eqnarray}
where the dot symbol represents the derivative with respect to time, $F_c$ represents the continuum luminosity received by 
an observer, and $\delta$ is the non-linear response parameter, which generally is a function of radius $r$ and time $t$. 
Here, we note that $\delta$ is closely related to the 
responsivity by $\eta = 1 + \delta$, which is formally defined as a derivative of the local line emission with respect to  
the incident hydrogen ionizing photon flux (see, e.g., \citealt{Korista2004,Goad2014}). We assume that $\delta$ is 
a constant temporally and spatially, as a result, 
\begin{equation}
\epsilon(r, t) \propto F_c^{1+\delta}(t-\tau).
\label{eqn_emissivity}
\end{equation}
Accordingly, the responsivity is also a constant temporally and spatially. 

\item All clouds rotate ``coherently'' around the central SMBH. Here we use ``coherently'' in the sense that 
each cloud's angular momentum is diverse but overall has a component oriented toward a common direction, i.e., the global
rotation axis of the BLR. Such coherent rotation is crucial for SA signals because 
orbital motion with fully random angular momentum would result in null SA signals (e.g., \citetalias{Gravity2018}; \citealt{Songsheng2019}). 
\end{itemize}

Below we first introduce mathematical preliminaries for RM and SA analysis
and then present our BLR dynamical models to calculate RM and SA signals.

\subsection{Mathematical Preliminaries}
\label{sec_math}
\subsubsection{Reverberation Mapping}
Under the scenario that BLRs are composed of discrete clouds, the total line emissions are given by summing up 
all clouds' emissions
\begin{equation}
F_l(v, t) = \sum_i \delta_D(v-v_i)\epsilon_i(t) = \sum_i \delta_D(v-v_i) A_i F_c^{1+\delta}(t-\tau_i),
\end{equation}
where $\delta_D(x)$ is the Dirac Delta function, $A_i$ is the response coefficient, and $v_i$ is the LOS velocity of $i$-th cloud. 
Because $\delta$ is assumed to be a constant independent on clouds' location, we can rewrite the above equation into 
a general form as (e.g., \citealt{Blandford1982, Peterson1993})
\begin{equation}
F_l(v, t) = \int \Psi_e(v, \tau) F_c^{1+\delta}(t-\tau)d\tau,
\label{eqn_rm}
\end{equation}
where the transfer function $\Psi_e(v, \tau)$ is defined by
\begin{equation}
\Psi_e(v, \tau) =  \sum_i \delta_D(v-v_i)\delta_D(\tau-\tau_i) A_i,
\label{eqn_tfe}
\end{equation}
which measures the response of BLRs at LOS velocity $v$ and time delay $\tau$ and 
the subscript ``$e$'' means  emissivity weighting, to be distinguished from responsivity weighting defined below.

For small continuum variations, we can use the first-order Taylor expansions to approximate each cloud's emissions as
\begin{eqnarray}
\epsilon_i(t) &=& A_i \left[\bar F_c + \Delta F_c(t-\tau_i)\right]^{1+\delta}\nonumber\\
&\approx& A_i \bar F_c + (1+\delta) A_i \Delta F_c(t-\tau_i),
\end{eqnarray}
where $\bar F_c$ represents the mean continuum flux and $\Delta F_c$ represents the variations. By defining 
\begin{equation}
\Delta F_l(v, t) = F_l(v, t) - \bar F_l(v),~~~\bar F_l(v) = \sum_i A_i\delta_D(v-v_i)  \bar F_c,
\end{equation}
we have 
\begin{eqnarray}
\Delta F_l(v, t) =  \int \Psi_r(v, \tau) \Delta F_c(t-\tau) d\tau,
\end{eqnarray}
where 
\begin{equation}
\Psi_r(v, \tau) = \sum_i (1+\delta)A_i\delta_D(v-v_i)\delta_D(\tau-\tau_i),
\label{eqn_tfr}
\end{equation}
and the subscript ``$r$'' means responsivity weighting.

Equations~({\ref{eqn_tfe}) and (\ref{eqn_tfr}) illustrate two branches of transfer functions widely used
in the literature. The major difference is that there is an extra responsivity factor $1+\delta$ in Equation~(\ref{eqn_tfr}). 
Since $\delta$ is a constant, emissivity-weighted and responsivity-weighted transfer functions are 
indeed identical. However, when $\delta$ is no longer constant spatially, these two transfer functions will be different 
and give rise to different centroid time lags (see also \citealt{Goad1993}).

\subsubsection{Spectroastrometry}
SA measures astrometry, namely, emissivity-weighted 
photocenters, as a function of LOS velocity $v$ at a specific time $t$ (\citealt{Beckers1982, Bailey1998}),
\begin{equation}
\boldsymbol{\varTheta}_l(v, t) = \frac{1}{D_{\rm A}}\frac{\sum_i \mathbi{x}_{i} \epsilon_i(t) \delta_D(v-v_i)}{\sum_i \epsilon_i(t) \delta_D(v-v_i)},
\label{eqn_tvt}
\end{equation}
where $\mathbi{x}_i$ is projected position of $i$-th cloud on the sky and $D_{\rm A}$ is the angular size distance of the source.
If we define the momentum of clouds' emissivity as 
\begin{equation}
\mathbi{M}_l(v, t) =\sum_i \mathbi{x}_{i} \epsilon_i(t) \delta_D(v-v_i)
\end{equation}
and the momentum transfer function as
\begin{equation}
\boldsymbol{\varPi}(v, \tau) = \sum_i \mathbi{x}_{i} \epsilon_i(t) \delta_D(v-v_i) \delta_D(\tau-\tau_i) A_i,
\end{equation}
we have reverberation mapping of the momentum
\begin{equation}
\mathbi{M}_l(v, t) = \int \boldsymbol{\varPi}(v, \tau) F_c^{1+\delta}(t-\tau)d\tau.
\label{eqn_momentum}
\end{equation}
The photocenters in Equation~(\ref{eqn_tvt}) can be rewritten into
\begin{equation}
\boldsymbol{\varTheta}_l(v, t)= \frac{1}{D_{\rm A}}\frac{\mathbi{M}_l(v, t) }{F_l(v, t)}.
\label{eqn_tvt2}
\end{equation}

In practice, there are two facts need to consider. First, the intensity we observed always consists of 
two components: one from the continuum and the other from the BLR. The continuum also contributes to the observed photocenters.
As a result, the observed photocenters are indeed given by
\begin{equation}
\boldsymbol{\varTheta}_{\rm all}(v, t) = \frac{F_c(t) \boldsymbol{\varTheta}_c(t) + F_l(v, t)\boldsymbol{\varTheta}_l(v, t)}{F_c(t) + F_l(v, t)}.
\end{equation}
It is reasonable to presume that the photocenters of the continuum do not change with wavelength/velocity. 
Therefore, we can use differential photocenters with respective to these of the continuum to further simplify calculations 
\begin{eqnarray}
\Delta \boldsymbol{\varTheta}_{\rm all}(v, t) &= &\boldsymbol{\varTheta}_{\rm all}(v, t) - \boldsymbol{\varTheta}_{c}(v, t) \nonumber \\
 &= & \frac{f(v, t)}{1 + f(v, t)}\Delta\boldsymbol{\varTheta}_l(v, t),
\end{eqnarray}
where $f(v, t)$ is the normalized line profile by the underlying continuum
\begin{equation}
f(v, t) = \frac{F_l(v, t)}{F_c(t)},
\end{equation}
and 
\begin{equation}
\Delta\boldsymbol{\varTheta}_l(v, t) = \boldsymbol{\varTheta}_l(v, t) - \boldsymbol{\varTheta}_c(v, t).
\end{equation}

Second, we usually measure astrometry only along a specific direction, e.g., a slit's spatial direction
through a spectrometer (e.g., \citealt{Stern2015}) or a baseline's direction through an interferometer (e.g., \citealt{Gravity2018}).
This effectively projects $\boldsymbol{\varTheta}_l(v, t)$ to a direction $\mathbi{j}$ as 
\begin{equation}
\theta_l(v, t) = \mathbi{j} \cdot \boldsymbol{\varTheta}_l(v, t).
\end{equation}
For interferometric observations, the observables are differential phases, related to photocenters as 
\begin{equation}
\Delta \phi(v, t) = -2\pi \frac{\mathbi{B}}{\lambda}\cdot \boldsymbol{\varTheta}_l(v, t),
\end{equation}
where $\mathbi{B}$ is the baseline and $\lambda$ is the wavelength, related to the velocity by $v/c = [\lambda/(1+z)-\lambda_0]/\lambda_0$,
where $c$ is the speed of light and $\lambda_0$ is the rest wavelength of the emission line.

In an extreme example where the continuum pulsas at time $t_0$, namely, $F^{1+\delta}_c(t) \sim \delta_{D}(t-t_0)$, we have the line emissions,
momentum transfer function, and photocenters as
\begin{eqnarray}
F_l(v, t) &=& \Psi_e(v, t-t_0),\\
\mathbi{M}_l(v, t) &=& \boldsymbol{\varPi}(v, t-t_0),\\
\Delta\boldsymbol{\varTheta}_{\rm all}(v, t) & = & \frac{1}{D_{\rm A}}\frac{f(v, t)}{1 + f(v, t)}\frac{\boldsymbol{\varPi}(v, t-t_0)}{\Psi_e(v, t-t_0)},
\end{eqnarray}
where we neglect the photocenters of the continuum.
Conversely, if the continuum flux is constant, the line emissions,
momentum transfer function, and photocenters are time independent accordingly, 
\begin{eqnarray}
F_l(v) &=& \Psi_e(v) \bar F_c,\label{eqn_flv}\\
\mathbi{M}_l(v) &=& \boldsymbol{\varPi}(v)\bar F_c,\\
\Delta\boldsymbol{\varTheta}_{\rm all}(v) & = & \frac{1}{D_{\rm A}}\frac{f(v)}{1 + f(v)}\frac{\boldsymbol{\varPi}(v)}{\Psi_e(v)},\label{eqn_tv}
\end{eqnarray}
where $\Psi_e(v)$ and $\boldsymbol{\varPi}(v)$ are delay integral of $\Psi_e(v, \tau)$ and $\boldsymbol{\varPi}(v, \tau)$, respectively.
Here, again, the photocenters of the continuum are neglected.

Equation~(\ref{eqn_tvt2}) implies that SA signals of BLRs are generally a function of time owing to responses of the emission line 
to the continuum variability. For 3C~273, the H$\beta$ light curve is almost invariable during the period (July, 2017-May, 2018) 
of SA observations. We expect a similar property for the Pa$\alpha$ line, therefore, we neglect time dependences of the 
SA signals and directly use Equations~(\ref{eqn_flv}-\ref{eqn_tv}) to fit the observed SA data.

\subsection{Broad-line Region Models}
We now proceed to construct phenomenological BLR dynamical models based on the framework of \cite{Pancoast2014} and \cite{Li2018}.
Section~\ref{sec_math} demonstrates that to calculate RM and SA signals, we need to specify geometry, dynamics, and
emissivity of BLR clouds.

\begin{figure}[t!]
\centering 
\includegraphics[width=0.47\textwidth]{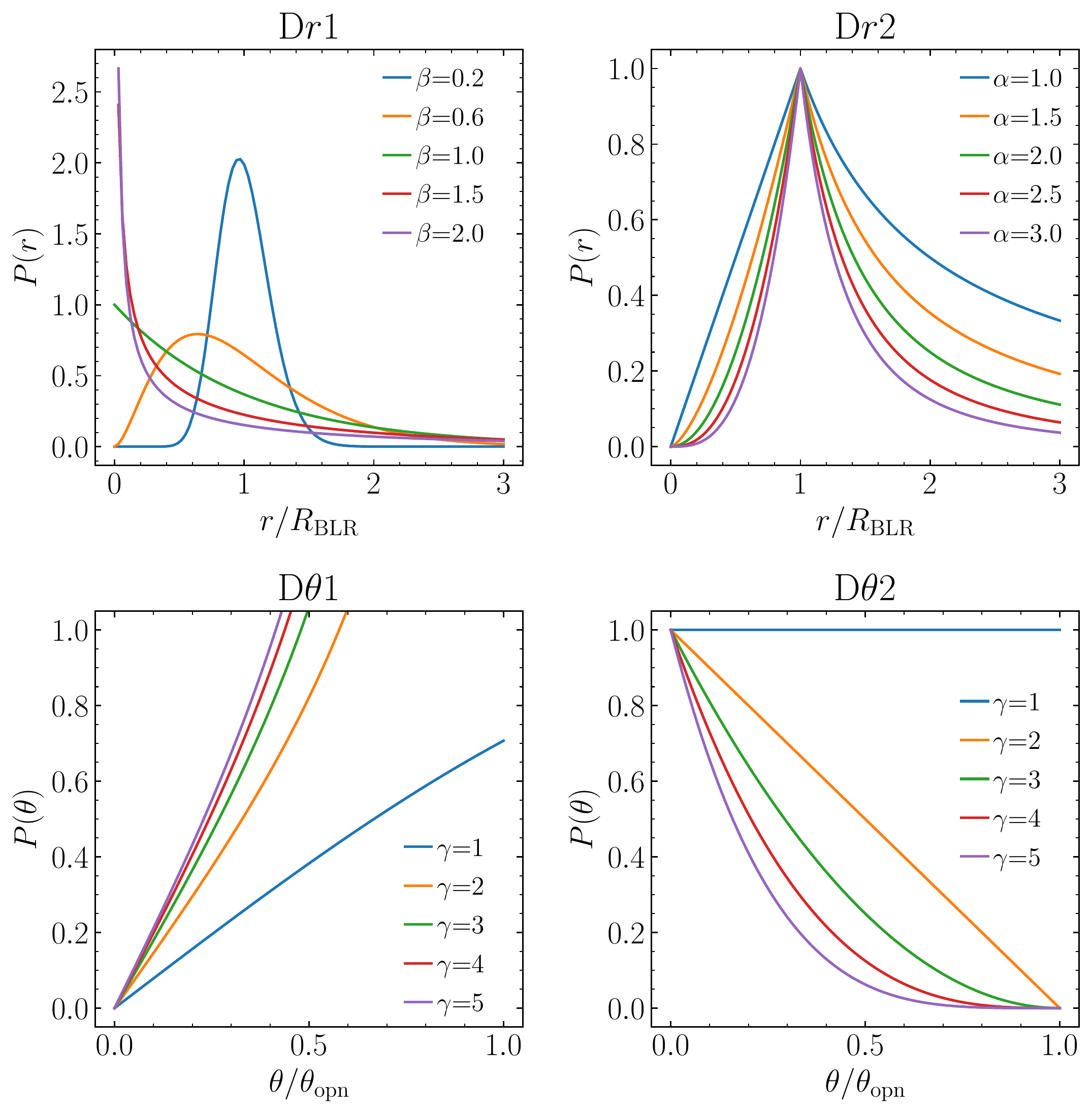}
\caption{Adopted distributions of BLR clouds along (top) radial and (bottom) vertical directions. See 
Section~\ref{sec_geometry} for the detail.}
\label{fig_dist}
\end{figure}

\subsubsection{Geometry} 
\label{sec_geometry}
Based on previous 2D RM observations (e.g., \citealt{Denney2010, Grier2013, Du2016, Du2018, Lu2016, DeRosa2018}; see also a review of 
\citealt{Peterson2014}), BLR clouds are assumed to follow a disk-like, axisymmetric geometry.
The BLR is viewed at an inclination angle $\theta_{\rm inc}$ and subtends an opening angle $\theta_{\rm opn}$.
Here, $\theta_{\rm inc}$ is defined to be the angle between the LOS and the symmetry axis of the BLR;
$\theta_{\rm opn}$ is defined as such that $\theta_{\rm opn}=90^\circ$ produces a spherical BLR while 
$\theta_{\rm opn}=0^\circ$ produces an infinitely thin disk-like BLR (see the schematic Figure 1 of \citealt{Li2013}).

In the radial direction, we adopt two types of distributions to assign clouds' distances from the central black hole, 
as shown in the top panels of Figure~\ref{fig_dist}.
\paragraph{\bf D$r$1} The distribution is parameterized by a gamma distribution (\citealt{Pancoast2014}). 
A BLR cloud is assigned a location by 
\begin{equation}
r = F_{\rm in} R_{\rm BLR} + (1-F_{\rm in}) \mathscr{R}, 
\label{eqn_gamma}
\end{equation}
where $F_{\rm in}$ is a fraction to denote the inner edge of the BLR and 
$\mathscr{R}$ is a random number generated from a gamma distribution
with a mean $R_{\rm BLR}$ and a standard deviation $\beta R_{\rm BLR}$. Here, the gamma distribution has a form of 
\begin{equation}
P(x|a, s) = \frac{x^{a-1}}{\Gamma(a)s^{a}}\exp\left(-\frac{x}{s}\right),
\end{equation}
where $a$ is the shape parameter, $s$ is the scale parameter, $\Gamma(a)$ is the Gamma function. 
In our parameterization, $a = 1/\beta^2$ and $s=\beta^2 R_{\rm BLR}$.

\paragraph{\bf D$r$2} The distribution is parameterized by a double power law (\citealt{Stern2015, Li2018}) as 
\begin{equation}
P(r) \propto \left\{ 
\begin{array}{ll}
\left(r/R_{\rm BLR}\right)^{\alpha} & {\rm for}~F_{\rm in}\leqslant r/R_{\rm BLR} \leqslant 1,\\
\left(r/R_{\rm BLR}\right)^{-\alpha} & {\rm for}~1\leqslant r/R_{\rm BLR} \leqslant F_{\rm out},
\end{array}
\right.
\label{eqn_powerlaw}
\end{equation}
where $\alpha$ is the slope of power law, and $F_{\rm in}$ and $F_{\rm out}$ are parameters that control
the inner and out radius.

In the $\theta$-direction, we also adopt two types of distributions as follows (see the bottom panels 
of Figure~\ref{fig_dist}).
\paragraph{\bf D$\theta$1} A distribution inclined to the outer faces of the BLR disk (\citealt{Pancoast2014})
\begin{equation}
\theta = \cos^{-1}\left[\cos\theta_{\rm opn} + (1-\cos\theta_{\rm opn})\times U^\gamma\right],
\end{equation}
where $\gamma$ is a free parameter to control to which extent clouds are clustered close to the 
outer faces and $U$ is a uniformly distributed random number between 0 and 1. 
This produces a distribution density 
\begin{equation}
P(\theta)\propto \sin\theta(\cos\theta-\cos\theta_{\rm opn})^{-(\gamma-1)/\gamma}.
\end{equation}
The case of $\gamma=1$ results in a uniform distribution of BLR clouds in terms of $\cos\theta$.

\paragraph{\bf D$\theta$2} A distribution inclined to the equatorial plane of the BLR disk
\begin{equation}
\theta = \theta_{\rm opn}\times (1 -U^{1/\gamma}),
\end{equation}
which produces a distribution density 
\begin{equation}
P(\theta)\propto(1-\theta/\theta_{\rm opn})^{\gamma-1}.
\label{eqn_dtheta2}
\end{equation}
The case of $\gamma=1$ gives a uniform distribution of BLR clouds in the $\theta$-direction.

We note that the above radial and vertical distributions are purely phenomenological.
There are not yet observations that directly constrain the most appropriate distributions.
We introduce these distributions to test for model dependence of our results.

\subsubsection{Dynamics} 
The prescription for BLR dynamics in \cite{Pancoast2014}'s phenomenological model is flexible to cover Keplerian motion,
inflows and outflows. We directly use this prescription and outline the essentials below for the sake of completeness.

In this prescription, a fraction $f_{\rm ellip}$ of clouds have bound elliptical Keplerian orbits and the remaining fraction $1-f_{\rm ellip}$
is in either inflowing ($0<f_{\rm flow}<0.5$) or outflowing ($0.5<f_{\rm flow}<1$).
Each cloud's velocities are first assigned in its orbital plane and then converted into three-dimensional velocities through appropriate 
coordinate transformations.
For elliptical orbits, the radial and tangential velocities are drawn from Gaussian distributions centered 
around a point $(v_r, v_\phi) = (0, v_{\rm circ})$ with standard deviations $\sigma_{\rho, \rm circ}$ and $\sigma_{\Theta, \rm circ}$, 
respectively. Here, $v_{\rm circ}=\sqrt{GM_\bullet/r}$ is the local Keplerian velocity. 
For inflowing/outflowing clouds, velocities are drawn similarly from Gaussian distributions centered 
around points $(v_r, v_\phi) = (\pm \sqrt{2} v_{\rm circ}, 0)$ with standard deviations $\sigma_{\rho, \rm rad}$ and $\sigma_{\Theta, \rm rad}$,
where ``$+$'' corresponds to outflows and ``$-$'' corresponds to inflows. It is straightforward to verify that 
a half of such generated inflows/outflows have bound orbits. To account for completely bound inflowing and outflowing orbits, 
the Gaussian distributions are allowed to rotate by an angle $\theta_e$ along the ellipse that has a semiminor axis $v_{\rm circ}$ 
in the $v_\phi$ direction and a semimajor axis $\sqrt{2}v_{\rm circ}$ in the $v_r$ direction (see Figure~2 in \citealt{Pancoast2014}).
Finally, all clouds' tangential velocities are forced to have the same sign (say, e.g., positive) so as to maintain
the overall coherent rotation (see Section~\ref{sec_assumptions}).

An extra small velocity offset is added to the LOS velocity to account for macroturbulences as 
\begin{equation}
v_{\rm turb} = \mathcal{N}(0, \sigma_{\rm turb}) v_{\rm circ},
\end{equation}
where $\mathcal{N}(0, \sigma_{\rm turb})$ is a random number generated from a normal distribution 
with a standard deviation of $\sigma_{\rm turb}$.

\subsubsection{Emissivity}
The clouds' emissivity is set according to Equation~(\ref{eqn_emissivity}).
To account for the possibility that BLR clouds are optically thick so that their emissions 
are anisotropic, we assign a weight to each cloud using a prescription (\citealt{Blandford1982})
\begin{equation}
 w = \frac{1}{2} + \kappa \cos\phi,
\label{eqn_kappa}
\end{equation}
where $\kappa$ is a parameter in the range $(-1/2, 1/2)$ and $\phi$ is the angle between the observer's and
cloud's LOS to the central black hole. A negative $\kappa$ corresponds to emissions 
preferentially from the far side of BLR clouds and a positive $\kappa$ corresponds to emissions 
preferentially from the near side. 

In addition, there is a possibility for some obscuring material in the equatorial plane that may not be
transparent to BLR clouds located below the equatorial plane. Following \cite{Pancoast2014}, we 
define a parameter $\xi$ to describe this effect. For $\xi=0$, the entire half of the BLR below the equatorial plane is completely 
obscured; while for $\xi=1$,  the equatorial plane becomes transparent so that the half can be directly observed.

\begin{figure}[t!]
\includegraphics[width=0.48\textwidth]{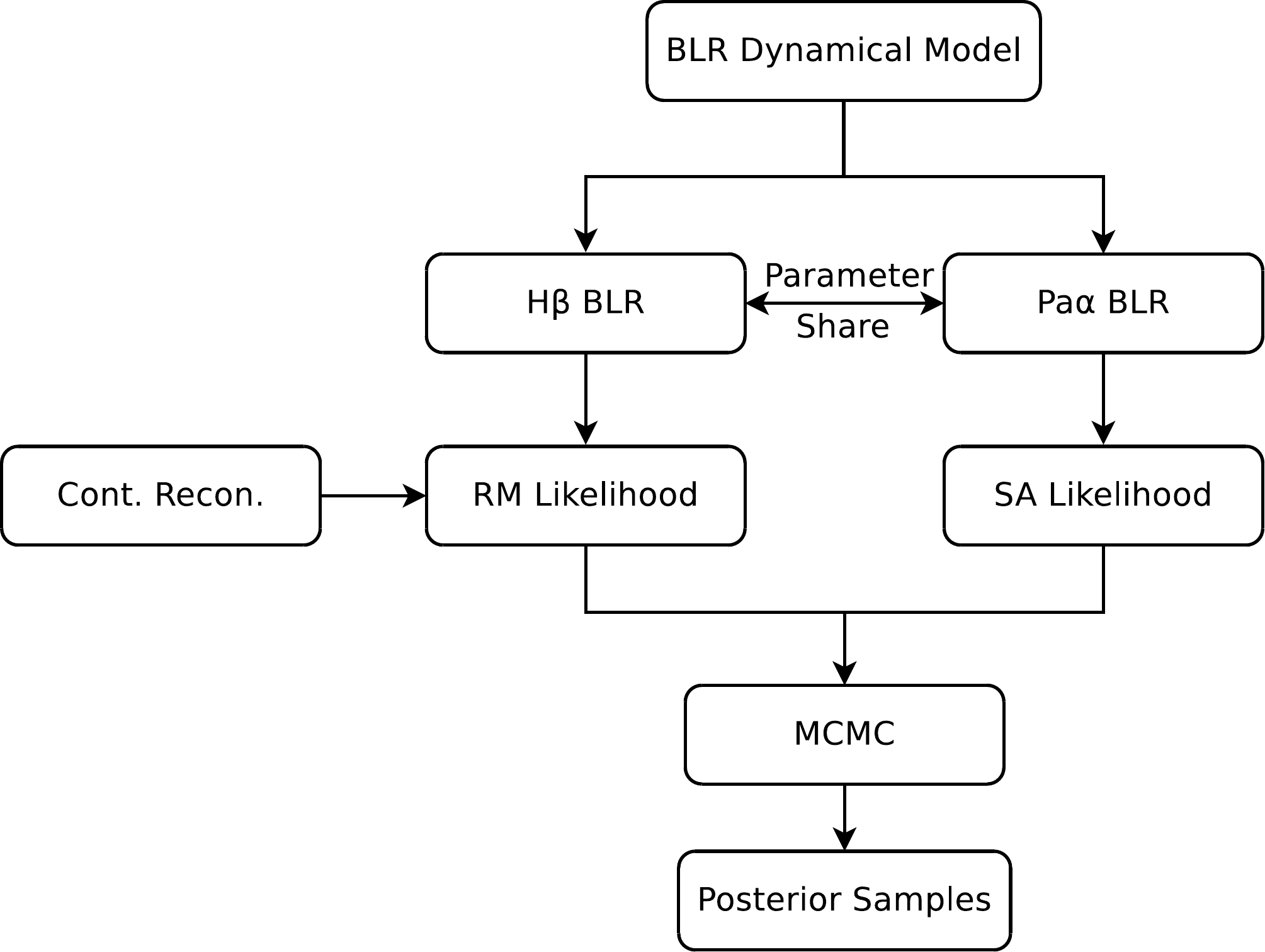}
\caption{A schematic diagram of the joint analysis of SA and RM observations for 3C~273.
The H$\beta$ BLR and Pa$\alpha$ BLR share inclination and black hole mass (see the text).
\label{fig_flowchart}}
\end{figure}

\subsection{Continuum Modeling and Long-term Detrending}\label{sec_cont}
We use the optical continuum light curve of 3C~273 compiled by \cite{Li2020}, which synthesized the data 
from the All-Sky Automated Survey for Supernovae project (\citealt{Kochanek2017}), 
the Small and Moderate Aperture Research Telescope 
System monitoring program (\citealt{Bonning2012}), {\it Swift} archives as well as from \cite{Zhang2019}.
We employ a damped random walk process to describe the continuum variability, which allows us to 
interpolate and extrapolate the observed continuum light curve at a given time grid (\citealt{Pancoast2014, Li2018}).
A damped random walk process is described by two parameters $\tau_{\rm d}$ and $\sigma_{\rm d}$, which 
represent the characteristic timescale of variations variation amplitude and the standard deviation 
of variations on long-timescale ($\gg \tau_{\rm d}$), respectively.
For an observed light curve, a reconstruction is obtained by starting with the best estimators at the 
given time grid and adding them Gaussian processes. The parameters $\tau_{\rm d}$ and $\sigma_{\rm d}$ 
together with the Gaussian processes are further constrained by fitting the emission line data
(see \citealt{Pancoast2014} and \citealt{Li2018} for the detail).

For 3C~273, an additional detrending treatment must be taken into account.
The continuum light curve  shows a distinct long-term trend that does not have 
a corresponding echo in the H$\beta$ light curve (\citealt{Zhang2019, Li2020}).
3C~273 is a flat-spectrum radio quasar with both comparably prominent jet and accretion disk emissions 
in its broad-band spectral energy distribution. As summarized in \cite{Li2020}, several pieces of evidence 
clearly demonstrated that the optical emissions of 3C~273 are composed of two components, one from the accretion disk 
and the other from the jet contribution. The most conclusive evidence come from the {\it Swift}
UV light curve of 3C 273, which is remarkably coincident with the scaled H$\beta$ light curve after correcting
for the time delay (see Figure~2 in \citealt{Li2020}) but distinct from the optical continuum 
light curve in the long-term trend. The jet contributions can naturally explain the non-echoed 
long-term trend in the continuum light curve.
\cite{Li2020} succeeded in detrending the continuum light curve using the radio light curve, which 
is generally consistent with the simple detrending using a linear polynomial.

In this work, we use the both approaches to detrend the continuum light curve.
For the approach of linear detrending, to keep the mean flux of the continuum light curve unchanged, 
we only need one free parameter to delineate the slope of the linear polynomial, namely,
\begin{equation}
F_c(t) = F'_{c}(t) - a(t-t_{\rm med}),
\end{equation}
where $F'_{c}(t)$ is the original continuum light curve, 
$a$ is the slope of the linear polynomial and $t_{\rm med}$ is the median time of the continuum light curve.
For the approach of detrending with a radio light curve, we use the radio data from the large-scale, fast-cadence
15 GHz monitoring program with the 40~m telescope at the Owens Valley Radio Observatory (\citealt{Richards2011}).
The detrended light curve is given by
\begin{equation}
F_c(t) = F'_{c}(t) - a_r\left[F_{r}(t-\tau_r) - b_r\right],
\label{eqn_radio}
\end{equation}
where $F_r$ is the radio light curve, $\tau_r$ is the time delay between the radio and continuum light curves
, and $a_r$ and $b_r$ are free parameters.
Because the radio light curve has intense sampling (nearly a daily cadence), we simply 
use the linear interpolation to derive the radio flux at a given time. 
We note that \cite{Li2020} adopted a more sophisticated detrending scheme, in which 
the radio light curve is regarded as a blurred echo of the optical light curve of the jet 
with a Gaussian transfer function. The inferred standard deviation of the Gaussian was about 13 days. 
It will significantly magnify the computation time to incorporate the scheme of \cite{Li2020} into present framework.
Therefore, we prefer to the above simple treatment in Equation~(\ref{eqn_radio}). Nevertheless, 
our treatment in Equation~(\ref{eqn_radio}) is equivalent to using an infinitesimally narrow Gaussian,
which is reasonable considering the quite narrow Gaussian width obtained by \cite{Li2020}.

\begin{deluxetable}{cccc}
\tablecolumns{4}
\tabletypesize{\footnotesize}
\tabcaption{\centering BLR Models with Combinations of Different 
Radial and Vertical Profiles of BLR Clouds and Continuum Detrending Approaches. \label{tab_param}}
\tablehead{
\colhead{Model} &  
\colhead{Radial Profile} &
\colhead{Vertical Profile} &
\colhead{Continuum Detrending}
}
\startdata
M1 & D$r1$  & D$\theta1$ & Linear \\
M2 & D$r1$  & D$\theta2$ & Linear \\
M3 & D$r2$  & D$\theta1$ & Linear \\
M4 & D$r2$  & D$\theta2$ & Linear \\
M5 & D$r1$  & D$\theta1$ & Radio \\
M6 & D$r1$  & D$\theta2$ & Radio \\
M7 & D$r2$  & D$\theta1$ & Radio \\
M8 & D$r2$  & D$\theta2$ & Radio \\
\enddata
\tablecomments{``Linear'' and ``radio'' denote long-term detrending of the continuum using a linear polynomial 
and the radio light curve, respectively (see Section~\ref{sec_cont}).}
\label{tab_models}
\end{deluxetable}

\begin{deluxetable*}{ccccl}
\tablecolumns{5}
\tabletypesize{\footnotesize}
\tabcaption{\centering Major Parameters for BLR Models. \label{tab_param}}
\tablehead{
\colhead{Parameter} &
\colhead{Prior} &
\colhead{Range} &
\colhead{Unit}   &
\colhead{Implication}
}
\startdata
$M_\bullet$        & LogUniform   & ($10^6, 10^{10}$)    & $M_\odot$ & Black hole mass\\
$D_{\rm A}$        & LogUniform   & (10, 5$\times10^3$)  & Mpc     & Angular-size distance\\
$\theta_{\rm inc}$ & CosUniform   & (0, 60)              & deg     & Inclination angle\\
PA                 & Uniform      & (-180, 180)          & deg     & Position angle on the sky (east of north)\\\hline
$R_{\rm BLR}$      & LogUniform   & (10, 400)            & lt-day  & Mean BLR radius (D$r$1)\\
$\beta$            & Uniform      & (0, 2)               & \nodata & Unit standard deviation of radial gamma distribution (D$r$1) \\
$F_{\rm in}$       & Uniform      & (0, 1)               & \nodata & Inner edge in units of $R_{\rm BLR}$ (D$r$1 and D$r$2)\\
$F_{\rm out}$      & LogUniform   & (1, 100)             & \nodata & Outer edge in units of $R_{\rm BLR}$ (D$r$2)\\
$\alpha$           & Uniform      & (1, 3)               & \nodata & Slope of the power-law radial distribution \\
$\theta_{\rm opn}$ & Uniform      & (0, 90)              & deg     & Opening angle\\
$\kappa$           & Uniform      & (-0.5, 0.5)          & \nodata & Anisotropy of the cloud emission\\
$\gamma$           & Uniform      & (1, 5)               & \nodata & Clustering of clouds in the vertical direction\\
$\xi$              & Uniform      & (0, 1)               & \nodata & Transparency of the equatorial material\\
$f_{\rm ellip}$    & Uniform      & (0, 1)               & \nodata   & Fraction of bound elliptical orbits\\
$f_{\rm flow}$     & Uniform      & (0, 1)               & \nodata   & Flag for determining inflowing or outflowing orbits\\\hline
\enddata
\tablecomments{``Uniform'' denotes a uniform prior, ``LogUniform'' denotes a uniform prior for the logarithm of the parameter,
and ``CosUniform'' denotes a uniform prior for the cosine of the parameter.}
\end{deluxetable*}

\subsection{A Suite of Models}

In Section~\ref{sec_geometry}, we introduce two radial distributions and two vertical distributions for BLR clouds.
Also, there are two approaches to detrend the continuum light curve (see Section~\ref{sec_cont}).
With these ingredients, we have 8 different combinations for BLR models, as summarized in
Table~\ref{tab_models}. In each model, the H$\beta$ and Pa$\alpha$ BLRs have the same parameterization and
share the inclination angle ($\theta_{\rm inc}$) and black hole mass ($M_\bullet$).
As such, we can test how the obtained results depend on specific BLR models
and justify the most probable model.
We note that those models (M1 and M5) with a combination of D$r$1 and D$\theta$1 are identical to 
\cite{Pancoast2014}'s BLR model.

In the calculations, we use $10^6$ particles to represent BLR clouds and set the maximum radius of BLR clouds fixed to 
800 lt-days, which is equal to the characteristic radius of the dusty torus in 3C~273 (\citealt{Gravity2020}).
Since the bulk of line emissions come from much smaller regions, this fixed outer radius does not influence the results.
Our following calculations also confirmed that the finally obtained mean BLR sizes are smaller than this outer radius.
This is as expected because the H$\beta$ time delay with respect to the continuum variations is around 150 days (\citealt{Zhang2019, Li2020})
and the Pa$\alpha$ BLR sizes also should not be much larger than the H$\beta$ BLR size in consideration of their line widths (see Figure~\ref{fig_compare}).

\section{Joint Bayesian Analysis}\label{sec_bayes}
\subsection{Bayesian Inference}
\label{sec_bayes_infer}
In Figure~\ref{fig_flowchart}, we show a schematic diagram of the joint analysis between SA and RM 
observation data for 3C~273.  Given a BLR dynamical model, we generate positions and velocities of clouds for the H$\beta$ and Pa$\alpha$ 
BLRs, which share the parameters of inclination and black hole mass. We then separately calculate simulated RM data (namely, a time series 
of H$\beta$ spectra) and SA data (including Pa$\alpha$ profile and differential phases). The simulated H$\beta$ spectra further need 
reconstructed continuum time series as a prior input.
By comparing against the observation data ($\mathbi{D}_{\rm RM}$, $\mathbi{D}_{\rm SA}$), we obtain the corresponding likelihoods
given a model parameter set $\boldsymbol\theta$.
The likelihood for RM data is given by 
\begin{equation}
P(\mathbi{D}_{\rm RM}|\boldsymbol{\theta}) = \prod_{ij}\frac{1}{\sqrt{2\pi}\sigma_{f, ij}}
\exp\left\{-\frac{[f_{ij}-f^{m}_{ij}(\boldsymbol{\theta}|F_c)]^2}{2\sigma_{f, ij}^2}\right\},
\end{equation}
where $f_{ij}$ and $\sigma_{f, ij}$ represent the observed flux density and uncertainty of the H$\beta$ line
at $i$-th epoch and $j$-th velocity bin, respectively, and $f^{m}_{ij}(\boldsymbol{\theta})$
represents the modeled flux density, which depends on the observed continuum light curve $F_c$.
The likelihood for SA data is given by
\begin{eqnarray}
P(\mathbi{D}_{\rm SA}|\boldsymbol{\theta}) &=& \prod_{i}\frac{1}{\sqrt{2\pi}\sigma_{F, i}}
\exp\left\{-\frac{[F_{i}-F^m_{i}(\boldsymbol{\theta})]^2}{2\sigma_{F, i}^2}\right\}\nonumber\\
&\times&\prod_{ij}\frac{1}{\sqrt{2\pi}\sigma_{\phi, ij}}
\exp\left\{-\frac{[\phi_{ij}-\phi^m_{ij}(\boldsymbol{\theta})]^2}{2\sigma_{\phi, ij}^2}\right\},
\end{eqnarray}
where $F_i$ and $\sigma_{F, i}$ represent the observed flux density and uncertainty of the Pa$\alpha$ line 
at the $i$-th velocity bin, $\phi_{ij}$ and $\sigma_{\phi, ij}$ represent the observed 
differential phase and uncertainty at $i$-th velocity bin and $j$-th baseline, $F^m_i$ and $\phi^{m}_{ij}$
represent the modeled flux density and differential phase, respectively. 

Based on the Bayes' theorem, 
the posterior probability is 
\begin{equation}
P(\boldsymbol\theta|\mathbi{D}_{\rm SA}, \mathbi{D}_{\rm RM}) = 
\frac{P(\mathbi{D}_{\rm SA}|\boldsymbol{\theta})P(\mathbi{D}_{\rm RM}|\boldsymbol{\theta})P(\boldsymbol\theta)}{P(\mathbi{D}_{\rm SA}, \mathbi{D}_{\rm RM})},
\end{equation}
where $P(\boldsymbol\theta)$ is the prior of parameters and $P(\mathbi{D}_{\rm SA}, \mathbi{D}_{\rm RM})$ is the Bayesian evidence.
Table~\ref{tab_param} summarizes the major parameters of BLR models and their prior ranges.
We use the Markov-chain Monte Carlo (MCMC) technique along with the diffusive nested sampling algorithm (\citealt{Brewer2011}) 
to explore the posterior probability. The diffusive nested sampling algorithm calculates Bayesian evidence
inherently (\citealt{Sivia2006}) and performs well in high-dimensional and multimodal distributions (\citealt{Brewer2018}). 

Following \cite{Li2018}, to account for any possible unknown errors in spectral reduction and decomposition,
we include an extra parameter $s_{\rm RM}$ to magnify the errors of the H$\beta$ profile data as 
$\sigma^2 = \sigma_d^2 +s^2_{\rm RM}{\bar \sigma_d}^2$, where $\sigma_d$ is the data errors 
and $\bar \sigma_d$ is the averaged error of all the H$\beta$ profiles. We adopt a prior of 
$P(x)=1/(1+x)$ for $s_{\rm RM}$, which behaves like a uniform prior when $s_{\rm RM}\ll1$ and like a logarithm prior 
when $s_{\rm RM}\gg1$ (\citealt{Gregory2011}). The prior range is set to (0, 10).

We incorporate the above joint analysis procedure into our previous developed package \texttt{BRAINS} (\citealt{Li2018}), 
which is publicly available at \url{https://github.com/LiyrAstroph/BRAINS}. 
The package was written in C language and runs on parallel computer clusters with the standardized message passing interface
based on a diffusive nested sampling library \texttt{CDNest} (\citealt{Li2020cdnest}).
For each model, we implement the sampling algorithm to generate 40000 ``raw'' parameter samples, from which
the final posterior samples are derived based on the recorded sampling information (such as the likelihood levels and associated 
prior mass). The convergence of the sampling is checked by inspecting the  
distribution of posterior weights as a function of prior mass.
To be specific, good sampling will see a significant peak in the distribution, which implies that 
the important parameter space where posterior weights get maximum is appropriately captured (see \citealt{Brewer2011} for a detail).
A posterior temperature ($T$) is introduced to anneal the likelihoods when deriving the posterior samples
(see also \citealt{Pancoast2014}). We use temperatures of 5-15 for the 8 models.
We finally determine the best estimates of parameters and their uncertainties 
by the medians of the corresponding posterior distributions and the 68.3\% confidence intervals, respectively. 

\begin{deluxetable*}{ccccccccc}
\renewcommand\arraystretch{1.2}
\tablecolumns{9}
\tabletypesize{\footnotesize}
\tabcaption{\centering Inferred Model Parameters. \label{tab_value}}
\tablehead{
\colhead{Model} &
\colhead{M1} &
\colhead{M2} &
\colhead{M3} &
\colhead{M4} &
\colhead{M5} &
\colhead{M6} &
\colhead{M7} &
\colhead{M8}
}
\startdata
$\log(M_\bullet/10^8M_\odot)$     & $0.10_{-0.33}^{+0.40}$ & $0.46_{-0.28}^{+0.51}$ & $0.21_{-0.26}^{+0.24}$ & $1.06_{-0.27}^{+0.21}$ & $-0.04_{-0.29}^{+0.25}$ & $0.46_{-0.57}^{+0.53}$ & $0.16_{-0.25}^{+0.13}$ & $0.79_{-0.48}^{+0.36}$ \\
$\log(D_{\rm A}/{\rm Mpc})$       & $3.09_{-0.23}^{+0.23}$ & $2.61_{-0.30}^{+0.39}$ & $3.14_{-0.22}^{+0.17}$ & $2.83_{-0.28}^{+0.32}$ & $2.94_{-0.31}^{+0.29}$ & $2.71_{-0.39}^{+0.35}$ & $3.12_{-0.19}^{+0.18}$ & $2.62_{-0.30}^{+0.34}$ \\
$\theta_{\rm inc}$                & $30_{-11}^{+10}$ & $8_{-3}^{+1}$ & $29_{-11}^{+7}$ & $5_{-1}^{+1}$ & $31_{-11}^{+9}$ & $9_{-3}^{+6}$ & $32_{-9}^{+6}$ & $7_{-2}^{+2}$ \\
PA/deg                            & $192_{-27}^{+20}$ & $190_{-41}^{+32}$ & $196_{-24}^{+18}$ & $196_{-36}^{+32}$ & $193_{-24}^{+22}$ & $197_{-36}^{+22}$ & $196_{-30}^{+19}$ & $202_{-36}^{+20}$ \\
\cutinhead{H$\beta$}
$\log(R_{\rm BLR}/\text{lt-day})$ & $2.10_{-0.23}^{+0.14}$ & $2.07_{-0.12}^{+0.22}$ & $1.15_{-0.10}^{+0.14}$ & $1.83_{-0.16}^{+0.43}$ & $1.93_{-0.23}^{+0.21}$ & $1.92_{-0.22}^{+0.34}$ & $1.15_{-0.13}^{+0.12}$ & $1.60_{-0.40}^{+0.35}$ \\
$\beta$                           & $1.62_{-0.05}^{+0.10}$ & $1.58_{-0.33}^{+0.16}$ & \nodata& \nodata& $1.63_{-0.05}^{+0.10}$ & $1.74_{-0.53}^{+0.13}$ & \nodata& \nodata\\
$F_{\rm in}$                      & $0.10_{-0.01}^{+0.01}$ & $0.14_{-0.04}^{+0.07}$ & $0.36_{-0.26}^{+0.39}$ & $0.34_{-0.26}^{+0.40}$ & $0.10_{-0.01}^{+0.01}$ & $0.11_{-0.02}^{+0.12}$ & $0.28_{-0.22}^{+0.22}$ & $0.41_{-0.28}^{+0.42}$ \\
$\log F_{\rm out}$                & \nodata& \nodata& $1.64_{-0.04}^{+0.06}$ & $0.98_{-0.79}^{+0.14}$ & \nodata& \nodata& $1.63_{-0.05}^{+0.03}$ & $1.20_{-0.27}^{+0.31}$ \\
$\alpha$                          & \nodata& \nodata& $1.21_{-0.03}^{+0.08}$ & $1.67_{-0.54}^{+0.33}$ & \nodata& \nodata& $1.21_{-0.03}^{+0.04}$ & $1.58_{-0.26}^{+0.37}$ \\
$\theta_{\rm opn}$                & $35_{-14}^{+11}$ & $75_{-26}^{+11}$ & $31_{-9}^{+8}$ & $78_{-13}^{+4}$ & $34_{-10}^{+11}$ & $72_{-21}^{+11}$ & $36_{-10}^{+6}$ & $77_{-12}^{+11}$ \\
$\kappa$                          & $-0.13_{-0.25}^{+0.35}$ & $0.07_{-0.35}^{+0.14}$ & $0.10_{-0.33}^{+0.31}$ & $-0.11_{-0.28}^{+0.42}$ & $0.06_{-0.47}^{+0.25}$ & $-0.21_{-0.25}^{+0.52}$ & $0.23_{-0.24}^{+0.18}$ & $0.05_{-0.46}^{+0.30}$ \\
$\gamma$                          & $3.85_{-1.69}^{+0.95}$ & $1.50_{-0.38}^{+0.76}$ & $4.54_{-1.10}^{+0.36}$ & $2.66_{-0.42}^{+0.65}$ & $3.70_{-1.59}^{+1.00}$ & $2.11_{-0.94}^{+1.30}$ & $4.62_{-0.67}^{+0.26}$ & $2.09_{-0.74}^{+0.99}$ \\
$\xi$                             & $0.63_{-0.32}^{+0.30}$ & $0.68_{-0.36}^{+0.21}$ & $0.64_{-0.18}^{+0.26}$ & $0.91_{-0.64}^{+0.07}$ & $0.59_{-0.27}^{+0.34}$ & $0.47_{-0.34}^{+0.37}$ & $0.65_{-0.30}^{+0.28}$ & $0.62_{-0.41}^{+0.34}$ \\
$f_{\rm ellip}$                   & $0.29_{-0.19}^{+0.19}$ & $0.29_{-0.24}^{+0.24}$ & $0.23_{-0.14}^{+0.18}$ & $0.27_{-0.17}^{+0.30}$ & $0.27_{-0.20}^{+0.21}$ & $0.39_{-0.29}^{+0.35}$ & $0.28_{-0.20}^{+0.22}$ & $0.47_{-0.25}^{+0.13}$ \\
$f_{\rm flow}$                    & $0.30_{-0.20}^{+0.31}$ & $0.28_{-0.18}^{+0.21}$ & $0.29_{-0.21}^{+0.18}$ & $0.31_{-0.20}^{+0.26}$ & $0.32_{-0.21}^{+0.22}$ & $0.32_{-0.25}^{+0.30}$ & $0.24_{-0.16}^{+0.18}$ & $0.34_{-0.27}^{+0.39}$ \\
$\delta$                          & $-0.20_{-0.05}^{+0.05}$ & $-0.19_{-0.04}^{+0.05}$ & $-0.20_{-0.03}^{+0.03}$ & $-0.14_{-0.06}^{+0.05}$ & $-0.54_{-0.02}^{+0.02}$ & $-0.54_{-0.02}^{+0.03}$ & $-0.53_{-0.02}^{+0.02}$ & $-0.51_{-0.03}^{+0.03}$ \\
\cutinhead{Pa$\alpha$}
$\log(R_{\rm BLR}/\text{lt-day})$ & $2.33_{-0.23}^{+0.18}$ & $1.86_{-0.22}^{+0.27}$ & $1.64_{-0.19}^{+0.16}$ & $1.94_{-0.32}^{+0.39}$ & $2.18_{-0.26}^{+0.24}$ & $2.01_{-0.37}^{+0.30}$ & $1.61_{-0.18}^{+0.12}$ & $1.74_{-0.37}^{+0.41}$ \\
$\beta$                           & $1.36_{-0.20}^{+0.19}$ & $0.98_{-0.69}^{+0.67}$ & \nodata& \nodata& $1.36_{-0.18}^{+0.20}$ & $1.23_{-0.67}^{+0.49}$ & \nodata& \nodata\\
$F_{\rm in}$                      & $0.13_{-0.04}^{+0.04}$ & $0.46_{-0.25}^{+0.32}$ & $0.45_{-0.34}^{+0.34}$ & $0.53_{-0.34}^{+0.33}$ & $0.13_{-0.04}^{+0.04}$ & $0.33_{-0.19}^{+0.36}$ & $0.50_{-0.32}^{+0.33}$ & $0.55_{-0.34}^{+0.34}$ \\
$\log F_{\rm out}$                & \nodata& \nodata& $1.54_{-0.32}^{+0.35}$ & $0.96_{-0.54}^{+0.69}$ & \nodata& \nodata& $1.54_{-0.32}^{+0.34}$ & $0.95_{-0.57}^{+0.55}$ \\
$\alpha$                          & \nodata& \nodata& $1.17_{-0.12}^{+0.17}$ & $2.22_{-0.69}^{+0.50}$ & \nodata& \nodata& $1.17_{-0.12}^{+0.13}$ & $2.16_{-0.70}^{+0.55}$ \\
$\theta_{\rm opn}$                & $32_{-13}^{+16}$ & $66_{-23}^{+17}$ & $33_{-9}^{+15}$ & $49_{-14}^{+20}$ & $35_{-13}^{+17}$ & $69_{-30}^{+15}$ & $34_{-9}^{+16}$ & $57_{-16}^{+18}$ \\
$\kappa$                          & $0.04_{-0.26}^{+0.25}$ & $-0.02_{-0.30}^{+0.35}$ & $0.10_{-0.34}^{+0.25}$ & $-0.01_{-0.32}^{+0.35}$ & $0.06_{-0.28}^{+0.28}$ & $0.02_{-0.33}^{+0.28}$ & $0.12_{-0.29}^{+0.22}$ & $0.03_{-0.31}^{+0.28}$ \\
$\gamma$                          & $2.73_{-1.17}^{+1.57}$ & $3.12_{-1.02}^{+1.14}$ & $2.65_{-1.18}^{+1.55}$ & $3.25_{-1.09}^{+1.21}$ & $2.75_{-1.24}^{+1.70}$ & $2.71_{-1.07}^{+1.18}$ & $3.02_{-1.50}^{+1.26}$ & $3.29_{-1.12}^{+0.94}$ \\
$\xi$                             & $0.63_{-0.35}^{+0.25}$ & $0.58_{-0.34}^{+0.32}$ & $0.60_{-0.31}^{+0.25}$ & $0.62_{-0.39}^{+0.26}$ & $0.66_{-0.42}^{+0.23}$ & $0.60_{-0.45}^{+0.28}$ & $0.67_{-0.38}^{+0.22}$ & $0.54_{-0.28}^{+0.32}$ \\
$f_{\rm ellip}$                   & $0.61_{-0.36}^{+0.28}$ & $0.59_{-0.35}^{+0.29}$ & $0.51_{-0.31}^{+0.34}$ & $0.61_{-0.39}^{+0.28}$ & $0.50_{-0.32}^{+0.34}$ & $0.62_{-0.35}^{+0.29}$ & $0.66_{-0.45}^{+0.24}$ & $0.59_{-0.41}^{+0.28}$ \\
$f_{\rm flow}$                    & $0.43_{-0.30}^{+0.35}$ & $0.45_{-0.31}^{+0.35}$ & $0.41_{-0.27}^{+0.34}$ & $0.57_{-0.40}^{+0.30}$ & $0.45_{-0.32}^{+0.36}$ & $0.49_{-0.29}^{+0.38}$ & $0.47_{-0.33}^{+0.35}$ & $0.58_{-0.42}^{+0.32}$ \\
\hline
$s_{\rm RM}$                      & 3.760 & 3.515   & 3.675  &  3.335  &  3.571  &3.685 & 3.540 & 3.532\\
$\chi^2_{\rm RM}/N_{\rm RM}$      & 0.985 & 0.986   & 0.985   & 0.986   & 1.039   & 0.987   & 1.034   & 1.012\\
$\chi^2_{\rm SA}/N_{\rm SA}$      & 1.304 & 1.326   & 1.378   & 1.401   & 1.344   & 1.309   & 1.329   & 1.317 \\
$\ln\mathcal{L}_{\rm max}$        & 0 & 61.71 & 17.15 & 105.45 & 14.34 & 17.65 & 26.23 & 42.12 \\
$\ln K$                           & 0 & 31.09 & 27.36 & 83.79 & 39.47 & 47.81 & 53.80 & 73.73 
\enddata
\tablecomments{The maximum likelihood $\ln\mathcal{L}_{\rm max}$ and Bayes factor $\ln K$ are given with respect to model M1. 
The Bayes factor $\ln K$ and $\ln\mathcal{L}_{\rm max}$ are given using a posterior annealing temperature of $T=8$ in the diffusive 
nested sampling algorithm. The values of $s_{\rm RM}$, $\chi^2_{\rm RM}$, and $\chi^2_{\rm SA}$ are these 
that maximize the likelihood ($\mathcal{L}_{\rm max}$).  $N_{\rm RM}=9450$ and $N_{\rm SA}=1000$ are the numbers of RM and SA data points, respectively.}
\end{deluxetable*}

\subsection{Model Selection}\label{sec_selection}
Among the 8 BLR models, we employ the Bayes factor to select the most probable one. 
The Bayes factor of two models, 
e.g., say M1 and M2, is defined by the ratio of their posterior probabilities and quantifies the relative merits 
of the two models. As usual, we assign equal priors for the two models. In this case, the Bayes factor 
is equal to the ratio of the Bayesian evidence (e.g., \citealt{Sivia2006})
\begin{equation}
K = \frac{P(M2|\mathbi{D}_{\rm SA}, \mathbi{D}_{\rm RM})}{P(M1|\mathbi{D}_{\rm SA}, \mathbi{D}_{\rm RM})} = 
\frac{P(\mathbi{D}_{\rm SA}, \mathbi{D}_{\rm RM}|M2)}{P(\mathbi{D}_{\rm SA}, \mathbi{D}_{\rm RM}|M1)}.
\end{equation}
A Bayes factor $K>1$ means that M2 is relatively preferable. A conventional threshold for a decisive preference is 
$K>100$ (\citealt{Jeffreys2003}).

There are otherwise independent observations that can aid us in model selection. First, 3C~273 displays a prominent 
radio jet (e.g., \citealt{Courvoisier1998}). The viewing inclination measured through superluminal motion ranges from 
3.8$^\circ$ to $\sim$15$^\circ$ (\citealt{Lobanov2001, Savolainen2006, Meyer2016, Jorstad2017}). This provides a useful constraint 
if assuming an alignment between the rotation axis of the BLR and the jet. Second, there are well-established correlations 
between black hole mass and luminosity/mass of classical bulges and ellipticals (see, e.g., \citealt{Kormendy2013}).
Using the {\it Hubble Space Telescope}/WFC3 images, \cite{Zhang2019} derived the bulge mass 
$\log\ (M_{\rm bulge}/M_\odot)=11.3\pm0.7$ and $K$-band bulge 
luminosity\footnote{Note that here, we use a cosmology of $H_0=70.5~{\rm km~s^{-1}~Mpc^{-1}}$ for consistency  
with \cite{Kormendy2013}.} $\log\ (L_{K, \rm bulge}/L_{K\odot}) = 11.68\pm0.28$, where $L_{K\odot}$ is the 
$K$-band luminosity of the Sun. We can compare our obtained black hole mass against these correlations to justify the most
probable model based on the Bayes factor.

\begin{figure*}[th!]
\centering 
\includegraphics[width=0.8\textwidth]{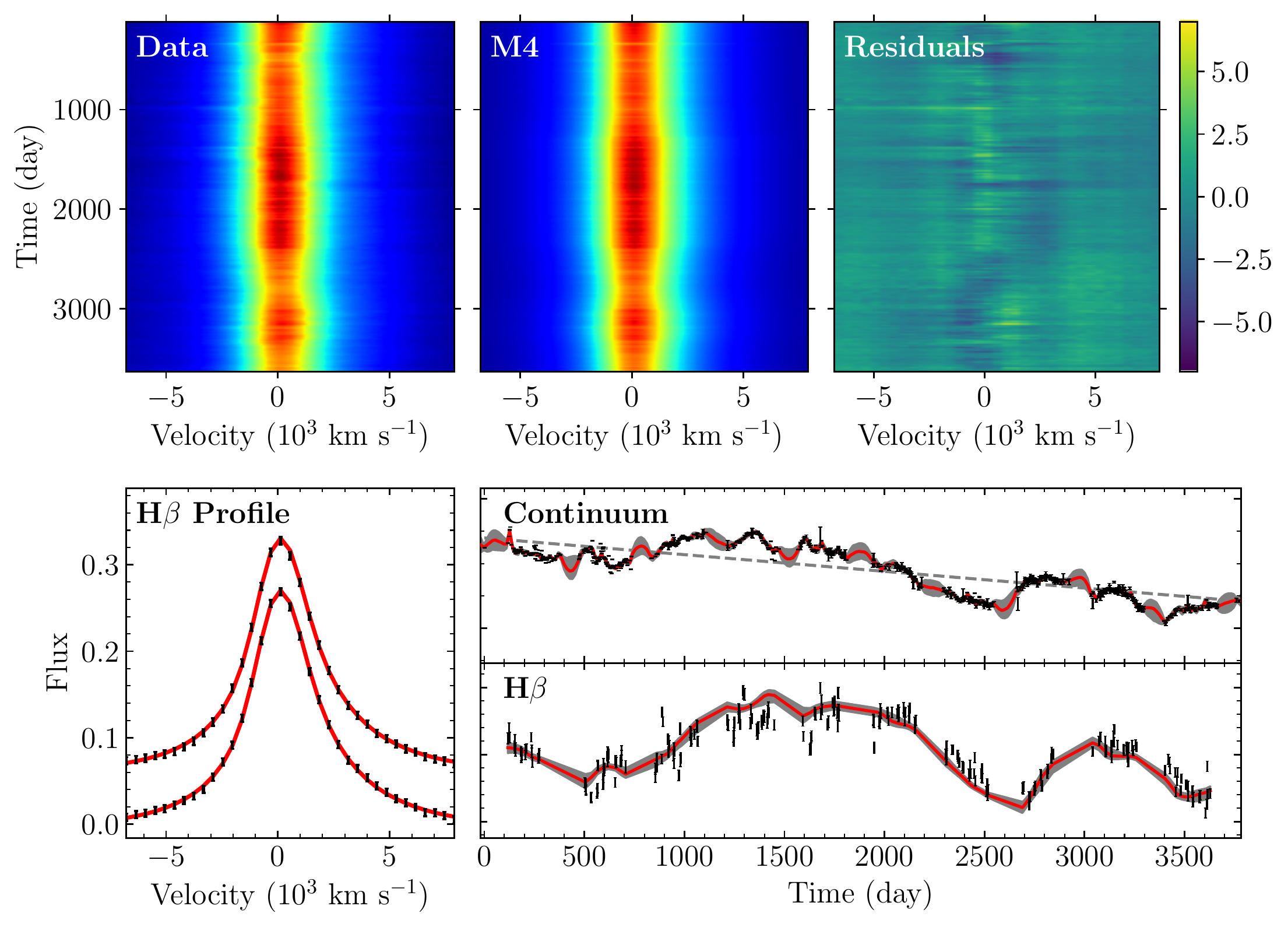}
\caption{Fits to the H$\beta$ RM data of 3C 273 using model M4. 
The top three panels show the observed H$\beta$ spectral time series, a model fit, and the standardized residuals 
between the observed data and model fit. The bottom left panel shows H$\beta$ profiles at two selected epochs, superposed 
upon the model fits with red solid lines. The bottom right panels show the time series of continuum and H$\beta$ fluxes. The red lines 
with grey shaded bands represent reconstructions from model fits. Grey dashed line presents the linear polynomial 
used to detrend the continuum light curve. The observations of continuum light curve are set to start at day zero.}
\label{fig_fits_rm}
\end{figure*}

\begin{figure*}[th!]
\centering 
\includegraphics[width=0.8\textwidth]{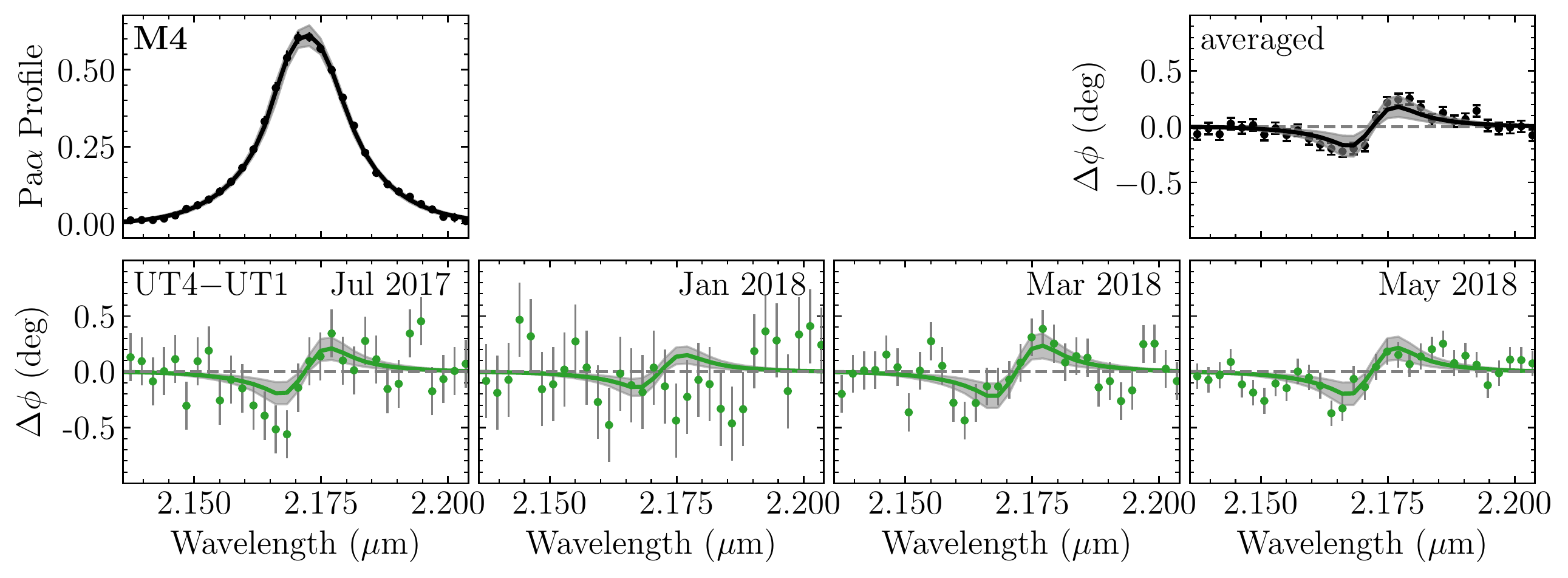}
\caption{Fits to the Pa$\alpha$ SA data of 3C~273 using model M4. The top left panel shows the time-averaged Pa$\alpha$ profile 
and the top right panel shows the time-averaged differential phase curves on three of the six baselines (UT4-3, UT4-2, and UT4-1; see also \citetalias{Gravity2018}).
The bottom panels show the differential phase curves on the UT4-1 baseline at four epochs observed by GRAVITY/VLTI
(Gravity Collaboration et al. 2018). The solid lines with grey shaded bands show reconstructions from model fits. The wavelengths are given
in observed frame. Fits to the full SA data (with six baselines at four epochs) are shown in Appendix. The wavelengths 
are given in observed frame.}
\label{fig_fits_sa_simp}
\end{figure*}

\section{Results}\label{sec_results}

\subsection{Overview}
\label{sec_overview}

In Table~\ref{tab_value}, we summarize the inferred values and uncertainties of major parameters 
for the 8 BLR models listed in Table~\ref{tab_models}. We also calculate the maximum likelihood 
$\ln \mathcal{L}_{\rm max}$ and the Bayes factor $\ln K$
for each model and tabulate their values with respect to model M1 in Table~\ref{tab_value}. 
The most probable model in light of both $\ln \mathcal{L}_{\rm max}$ and Bayes factor is M4, which has 
the maximum $\ln \mathcal{L}_{\rm max}$ and $\ln K$. In model M4, the continuum light curve is detrended by 
a linear polynomial, indicated by a grey dashed line in the bottom right panel of Figure~\ref{fig_fits_rm}. 
The radial profile of BLR clouds 
are described by a double power law (Equation~\ref{eqn_powerlaw}) with a slope parameter $\alpha\sim1.7$ 
for the H$\beta$ BLR and $\alpha\sim2.2$ for the Pa$\alpha$ BLR. In the vertical direction, BLR clouds 
are  more inclined to distribute around the equatorial plane (Equation~\ref{eqn_dtheta2}), although
the opening angles for both H$\beta$ and Pa$\alpha$ BLRs are large.

In Figures~\ref{fig_fits_rm}, we show fits to H$\beta$ RM data using model M4. 
The large-scale variation features of the H$\beta$ spectral time series are generally reproduced, 
although there are still minor patterns in the residuals.
For the integrated H$\beta$ fluxes in right bottommost panel of Figure~\ref{fig_fits_rm}, 
our fits again well match the large-scale variations, nevertheless,
there appear slight excesses around 1300 days and minor deficits around 2300-2500 days and 3100-3500 days.
The reasons for the above minor deviations might be twofold: first, the present simple 
dynamical BLR models are not expected to reproduce all the fine variation structures;
second, for simplicity, we used a linear polynomial to detrend the continuum light curve,
which can be only regarded as a zero-order approximation (\citealt{Li2020}).
In Figure~\ref{fig_fits_sa_simp}, we show fits to the Pa$\alpha$ profile and differential phase curves.
For clarity, we only plot the phase curves on one baseline (UT4-UT1) and the time-averaged phase curve 
on three of the six baselines (UT4-UT3, UT4-UT2, and UT4-UT1; see also \citetalias{Gravity2018}). Full fits 
to the phase curves on all six baselines are shown in Appendix. Overall,
the Pa$\alpha$ profile and differential phase curves are also 
well reproduced, in particular for the baselines UT4-UT3, UT4-UT2, and UT4-UT1 where 
the phase signals are of high quality.

\begin{figure*}[th!]
\centering 
\includegraphics[width=0.8\textwidth]{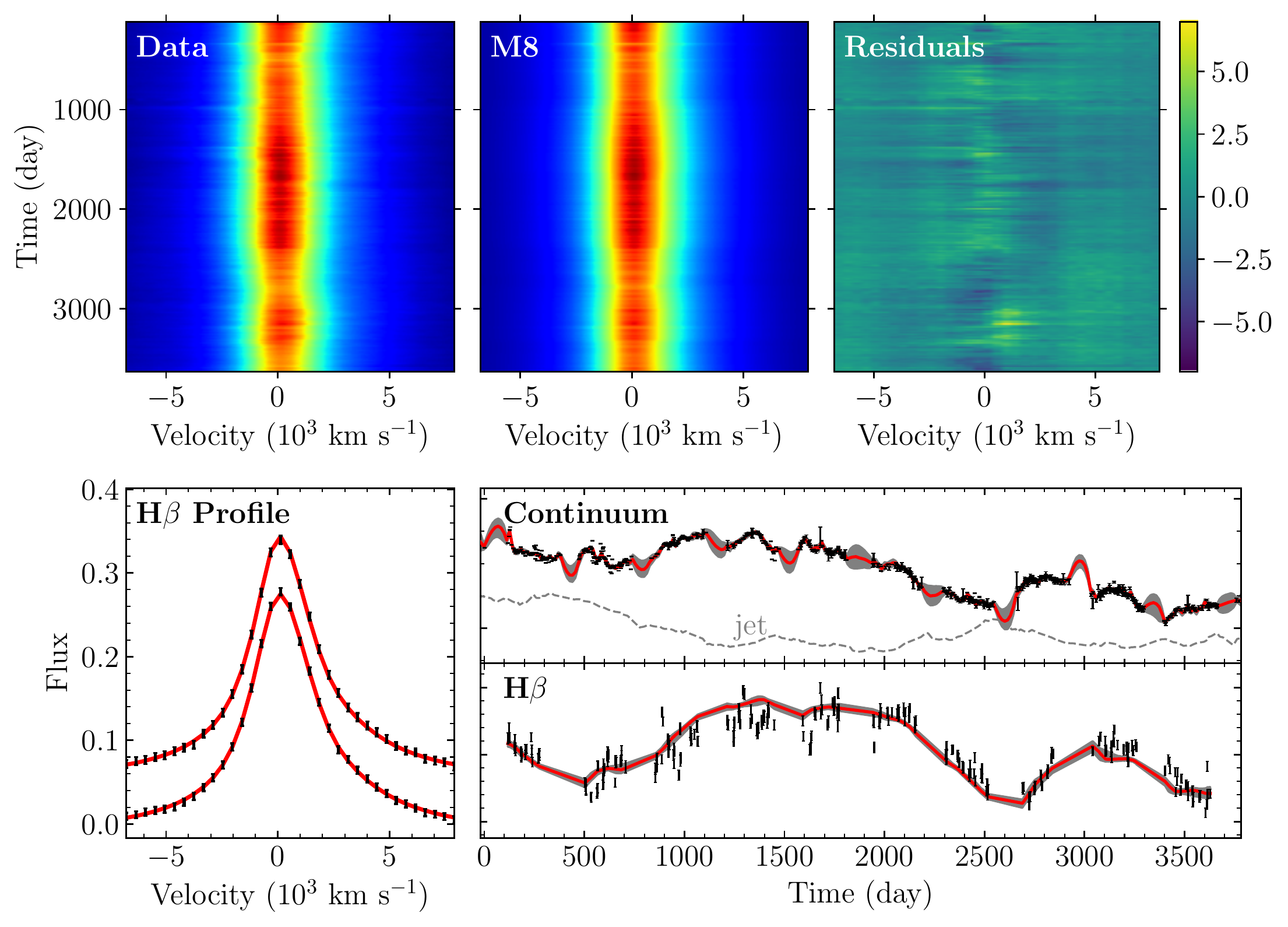}
\caption{Same as Figure~\ref{fig_fits_rm}, but using model M8. Grey dashed line represents the trend subtracted from the 
continuum light curve, which is derived from the 15 GHz radio light curve with shifting and scaling (see Equation~\ref{eqn_radio}).}
\label{fig_fits_rm_m8}
\end{figure*}

\begin{figure*}
\centering 
\includegraphics[width=0.8\textwidth]{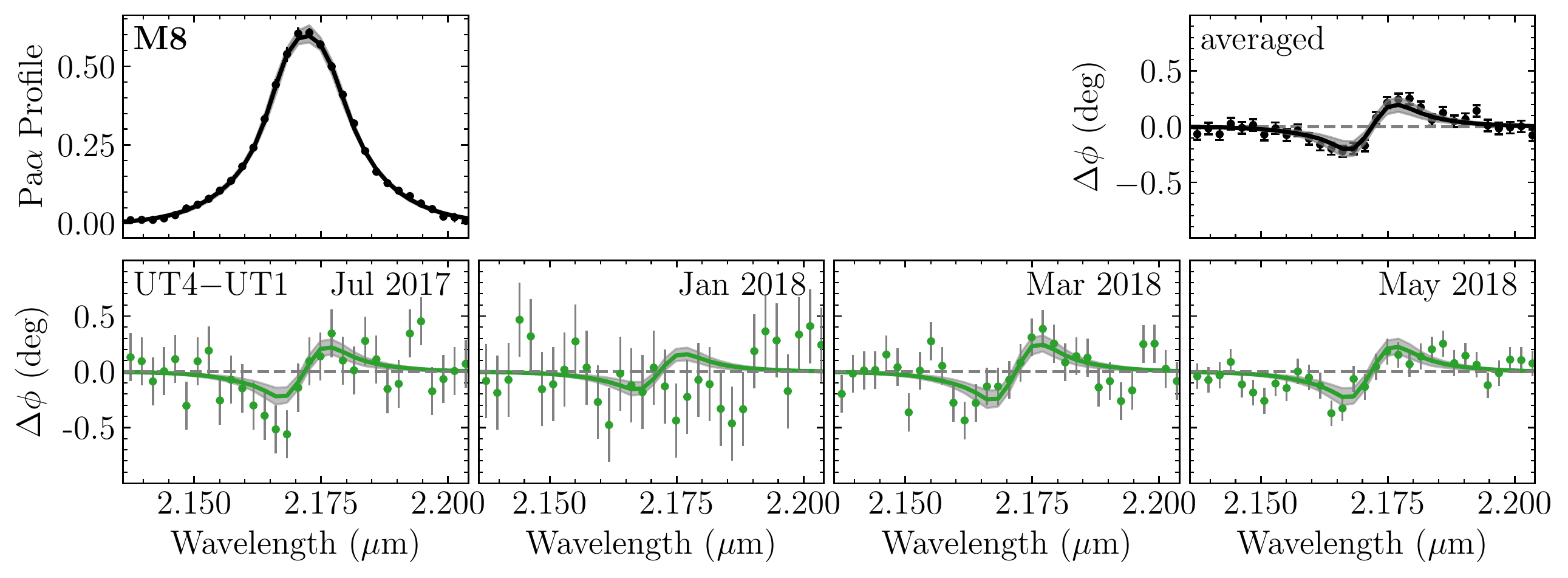}
\caption{Same as Figure~\ref{fig_fits_sa_simp} but using model M8. Fits to the full SA data (with six baselines at four epochs) are shown in Appendix.}
\label{fig_fits_sa_simp_m8}
\end{figure*}

For the sake of comparison, in Figures~\ref{fig_fits_rm_m8} and \ref{fig_fits_sa_simp_m8}, 
we show fits to the H$\beta$ RM data and Pa$\alpha$ SA data using model M8, which has the highest rank among 
the models (M5-M8) that detrend the continuum light curve with the 15 GHz radio light curve.
Again, full fits to the differential phase curves on six baselines with model M8 are shown in Appendix.
Similar to model M4, the large-scale H$\beta$ variation features and P$\alpha$ differential phase curves
are overall captured. However, compared to model 4, model 8 additionally yields relatively mild
deficits around 3300-3500 days in the H$\beta$ light curves. The fitting results 
with the other models are generally similar with M8 and therefore not shown.

Figure~\ref{fig_stats} compares the posterior values of black hole mass ($M_\bullet$), angular-size distance ($D_{\rm A}$),
inclination angle ($\theta_{\rm inc}$), and position angle (PA) for the 8 models.
A remarkable point is that the vertical distribution D$\theta$1 systematically yields a larger inclination angle
of $\sim 30^\circ$, whereas the distribution D$\theta$2 yields an inclination angle of $\sim 6^\circ$.
The latter value is consistent with the inclination $3.8^\circ$-$15^\circ$ of the jet in 3C~273 measured by 
radio observations (e.g., \citealt{Lobanov2001, Savolainen2006, Meyer2016, Jorstad2017}). 
The main reason for such a dichotomy is as follows (see also discussions 
in \citealt{Grier2017} and \citealt{Li2018}). In a case of a thin disk-like BLR with 
clouds preferentially distributed close to the equatorial plane (e.g., D$\theta$2), the resulting line profiles 
are always double-peaked for a large inclination angle. Therefore, the observed single-peaked profiles of both the 
H$\beta$ and Pa$\alpha$ lines require a small (nearly face on) inclination. By contrast, when BLR clouds preferentially 
cluster around the outer faces (e.g., D$\theta$1), the line profiles tend to be flat in the core as the opening angle 
increases. Meanwhile, to generate single-peaked profiles, the opening angle should be larger than the inclination angle.
As a result, the observed single-peaked profiles require the opening angle to be as small as possible, but still larger than
the inclination angle. The final resulting opening angle is approximately equal to the inclination angle, roughly about 30$^\circ$-40$^\circ$.
Such an approximate equality was commonly seen in previous studies of BLR dynamical modeling 
(e.g., \citealt{Pancoast2014b, Grier2017, Li2018, Williams2018}). In this case, the inclination angle 
cannot be as small as that in the case of D$\theta$2 distribution, otherwise 
the yielded transfer function will be too spiky in wavelength and time-delay space to fit the observations.

\begin{figure*}[th!]
\centering 
\includegraphics[width=0.7\textwidth]{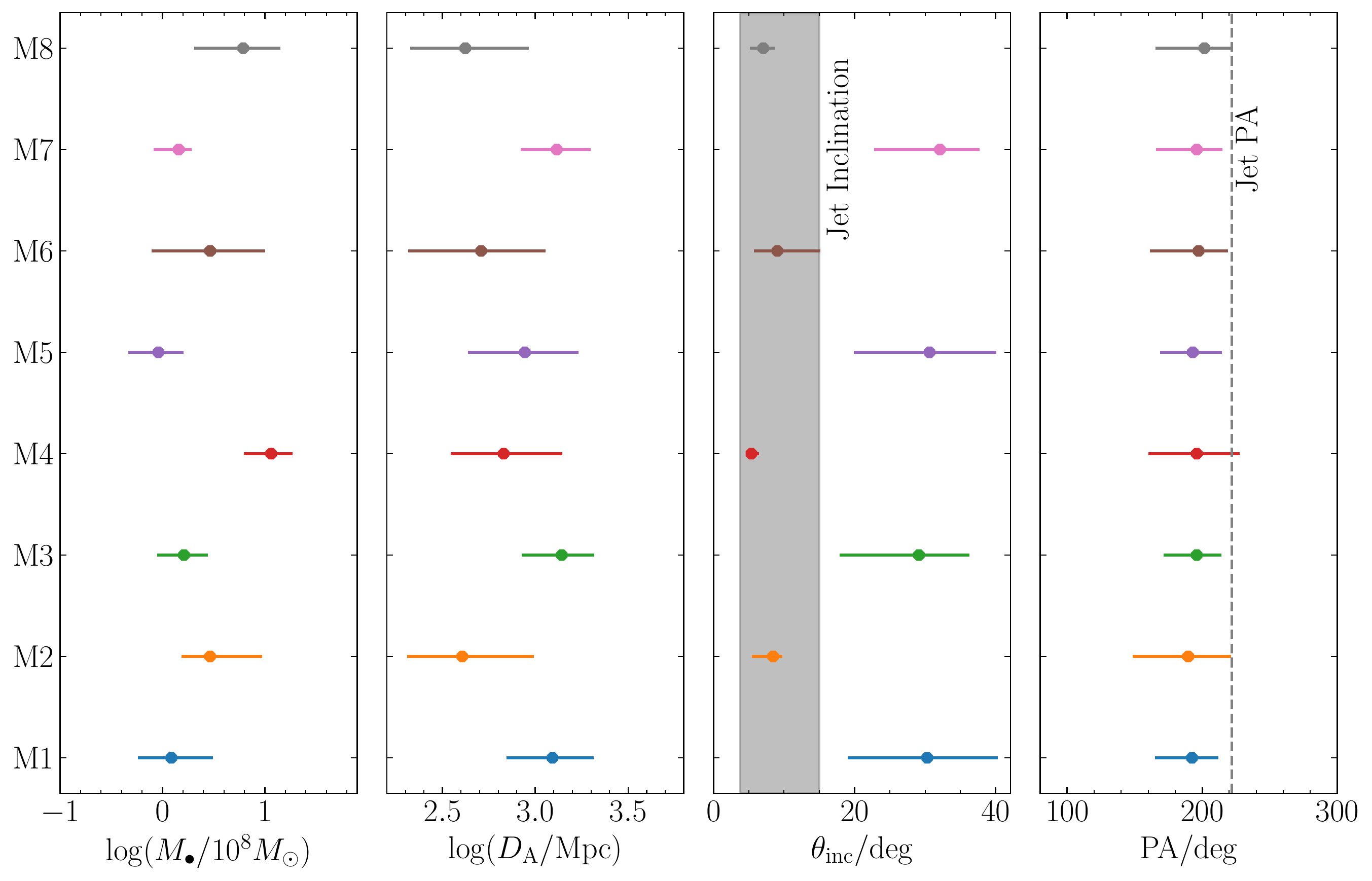}
\caption{Inferred black hole mass, angular-size distance, inclination angle, and position angle for different models. 
In the third panel, the vertical shaded band 
represents the inclination angle (3.8$^\circ$-15$^\circ$) of the jet in 3C 273 from radio measurements.
In the rightmost panel, the vertical dashed line represents the position angle of the jet.}
\label{fig_stats}
\end{figure*}

The PAs are overall similar, around 200$^\circ$ with a typical uncertainty of $\sim20^\circ$-$30^\circ$, 
insensitive to the radial and vertical distributions of BLR clouds. This is
because PA is mainly determined by the projected positions of the six baselines
on the sky. The PA of the large-scale jet of 3C~273 is about 222$^\circ$ (e.g., \citealt{Roeser1991}), marginally
coincident with that of the Pa$\alpha$ BLR. Figures~\ref{fig_stats} and \ref{fig_posts}
also illustrate that the detrending approaches (using a linear polynomial or the radio light curve) 
have limited influences on the final results.

\begin{figure}[th!]
\centering 
\includegraphics[width=0.48\textwidth]{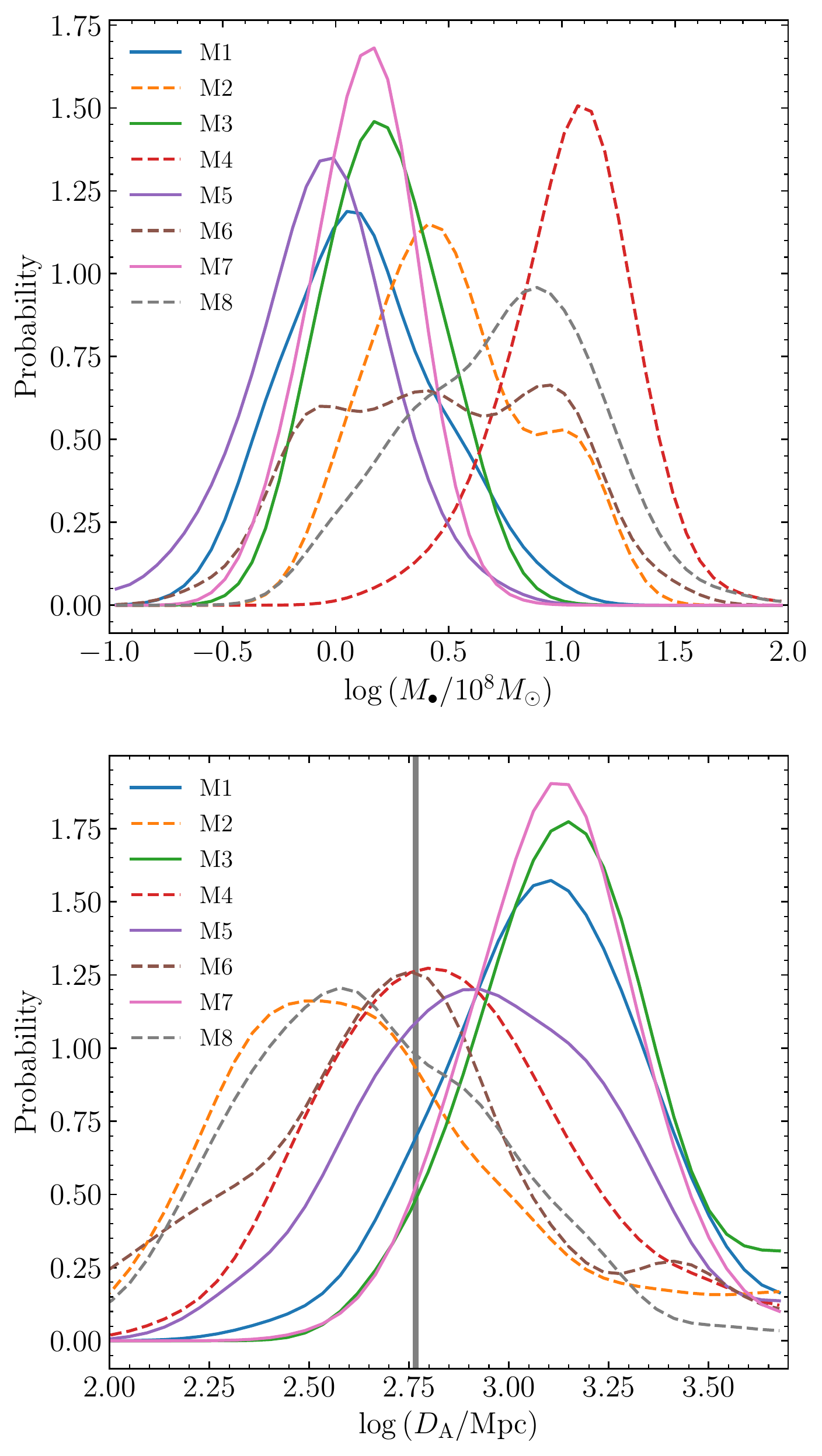}
\caption{Posterior distributions of black hole mass and angular-size distance.
Solid and dashed colored lines correspond to BLR models with the D$\theta$1 and D$\theta$2
distribution, respectively.
In the bottom panel, the vertical grey line represents an angular-size distance of $D_{\rm A}=585$~Mpc.}
\label{fig_posts}
\end{figure}

\subsection{The Black Hole Mass}
We calculate the best estimates and uncertainties of black hole mass and illustrate the posterior distributions
in the top panel of Figure~\ref{fig_posts} for the 8 models. 
The obtained black hole masses are concentrated around $10^8M_\odot$ for cases of vertical distributions D$\theta$1
and around $10^9M_\odot$ for cases of vertical distributions D$\theta$2.
Such a large difference is mainly caused by the inclination angles.
The LOS velocity is proportional to $\sin\theta_{\rm inc}$. 
Consequently, if regardless of other factors, 
the black hole mass depend on the inclination  as
\begin{equation}
M_\bullet \propto 1/\sin^{2}\theta_{\rm inc}.
\label{eqn_mass_inc}
\end{equation}
As described above, the obtained inclination for cases of  D$\theta$1 distribution is about 30$^\circ$ while 
the inclination for cases of D$\theta$2 distribution is about 6$^\circ$. This can easily lead to 
a difference in black hole mass by one order of magnitude.

The most probable model M4 yields a black hole mass of $\log (M_\bullet/M_\odot) = 9.06_{-0.21}^{+0.27}$.
In Figure~\ref{fig_mass_relation}, we show the locations of 3C~273 on the relations between 
black hole mass and bulge mass and $K$-band luminosity compiled by \cite{Kormendy2013}. 
Figure~\ref{fig_mass_relation} illustrates that the obtained black hole mass with model M4 is remarkably 
consistent with the relationships between black hole mass and bulge properties (within 1.7$\sigma$), despite a large 
uncertainty of the bulge mass of 3C~273. This reinforces the model selection based on the Bayes factor.
Moreover, the black hole masses of $\sim 10^8M_\odot$ obtained by models with the D$\theta1$ distribution
(M1, M3, M5, and M7) deviate at $3-5\sigma$ confidence levels from the relationships between black hole mass and bulge
properties and therefore are statistically not preferable.

\subsection{The Angular-size Distance}
Figures~\ref{fig_stats} and \ref{fig_posts} illustrate that  the values of $D_{\rm A}$ also 
depend on vertical distributions of BLR clouds.
In models with D$\theta$2 distributions (M2/4/6/8), the inferred $D_{\rm A}$ is generally 
larger than that in models with D$\theta$1 distributions (M1/3/5/7) by about 0.1-0.5 dex.
The $\Lambda$CDM cosmology with the latest {\it Planck} measurements (\citealt{Planck2020}) yields an angular-size 
distance of $\log\,(D_{\rm A}/{\rm Mpc})=2.77$ to 3C 273. If we use 
this value as the fiducial distance, the models with D$\theta$2 distributions yield 
more reasonable inferences. Among models with the D$\theta$1 distribution, 
M1 and M5 yield results in line with the $\Lambda$CDM cosmology
within the 1$\sigma$ uncertainty; the distances obtained by M3 and M7 are a bit larger by 0.4 dex, but still consistent with the 
fiducial value within a $2.5\sigma$ level.
Both models M3 and M7 employ the D$r$2 distribution (double power law) in radial direction 
and D$\theta$1 distribution (concentrated around the equatorial plane) in vertical distribution. 
The discrepancy possibly implies that a combination of D$r$2 and D$\theta$1 might not be preferred.

We stress that the present SARM analysis can naturally obtain the angular-size distance. 
The above comparison with the distance from the $\Lambda$CDM cosmology is mainly aimed to illustrate the differences among the 8
models.

\begin{figure*}[t!]
\centering 
\includegraphics[width=0.45\textwidth]{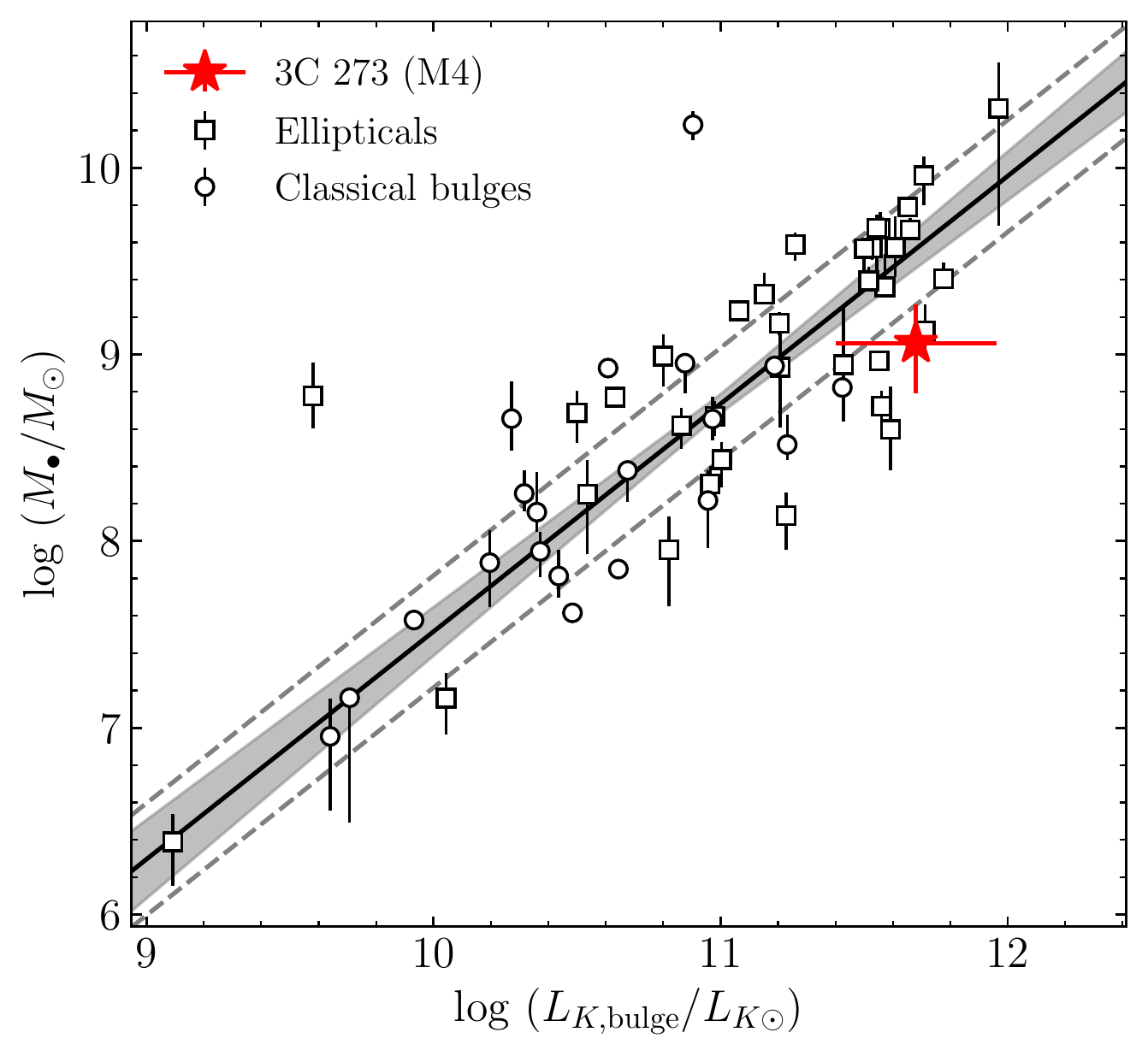}~~~
\includegraphics[width=0.45\textwidth]{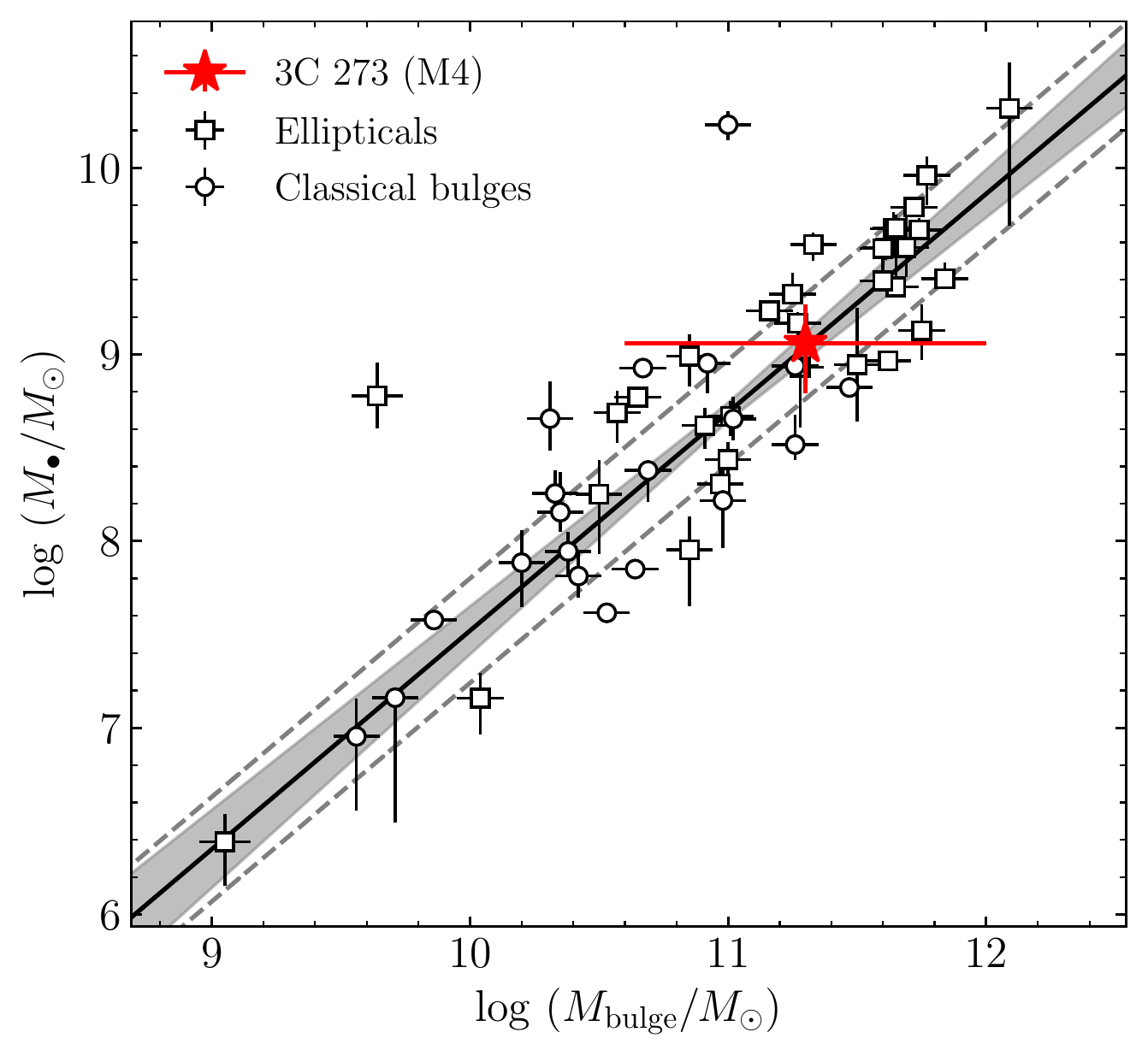}
\caption{The relations between black hole mass and (left) $K$-band bulge luminosity and (right) bulge mass  of classical 
bulges and ellipticals compiled by \cite{Kormendy2013}.
Solid lines with shaded areas represent the linear regression fits from \cite{Kormendy2013}. 
Grey dashed lines represent the intrinsic scatters of the relations.
Superimposed is 3C~273 (red star) with the mass obtained using model M4 and the bulge mass and luminosity 
derived from \cite{Zhang2019} using the {\it Hubble Space Telescope}/WFC3 images.}
\label{fig_mass_relation}
\end{figure*}

\subsection{Geometry and Kinematics of the H$\beta$ and Pa$\alpha$ BLRs}
In Table~\ref{tab_value}, we summarize values of the major parameters of all models for both the H$\beta$ and Pa$\alpha$ BLRs.
In Figure~\ref{fig_compare}, we compare the posterior distributions of major parameters of the most probable mode M4.
For all models (except M2), the obtained H$\beta$ BLR sizes are generally smaller than the Pa$\alpha$ BLR sizes, consistent 
with the observation that the H$\beta$ line has a broader width than the Pa$\alpha$ line (see Figure~\ref{fig_prof}). 
In model M2, the H$\beta$ BLR sizes are comparable to or even larger than the Pa$\alpha$ BLR sizes.
M2 uses D$\theta$2 distribution in the vertical direction (see Table~\ref{tab_models}). The parameter $\gamma$ is a bit larger 
for the Pa$\alpha$ line ($\gamma\sim3$) than that for the H$\beta$ line ($\gamma\sim1.5$).
Larger $\gamma$ means that the BLR clouds are more concentrated towards the equatorial plane,
giving rise to a larger SA phase amplitude owing to a decrement in the stochastic motion along the LOS. 
As a result, the Pa$\alpha$ BLR size is reduced to match the observed phase amplitudes (\citealt{Songsheng2021}). 
However, considering the larger $\beta$ value of the H$\beta$ BLR, the resulting clouds are still 
more compactly distributed than those of the Pa$\alpha$ BLR.

Both the H$\beta$ and Pa$\alpha$ BLRs have large opening angles ($\sim40^\circ$-$80^\circ$), indicating that 
they are overall geometrically thick. The values of parameter $\xi$ for all models are $\sim0.4-0.8$, meaning that 
the equatorial mediums are partially transparent. The values of parameter $\gamma$ systematically deviate from 
unity so that the vertical distributions of BLR clouds are neither uniform in light of $\cos\theta$ nor uniform 
in light of $\theta$ (see Section~\ref{sec_geometry}). The parameter $\kappa$, used to describe 
the anisotropy of BLR emissions (Equation~\ref{eqn_kappa}), has a large posterior range. In models M4 and M8, the posterior
distribution of $\kappa$ for the H$\beta$ BLR is double peaked around $\kappa=-0.5$ and $0.3$ 
(e.g., see Figure~\ref{fig_compare}). Overall, $\gamma$ is not well constrained under present 
data quality. 

In Figure~\ref{fig_clouds}, for the sake of comparison, we show examples of 
cloud distributions of the H$\beta$ and Pa$\alpha$ BLRs for models M1-M4.
We can clearly find that the H$\beta$ BLR is more compact that the Pa$\alpha$ BLR.
A recent photoionization calculation by \cite{Zhang2021} showed that the emissivity of the H$\beta$ line
decreases with radius more rapidly than that of the Pa$\alpha$ line. This indicates that 
the difference between the two BLRs arises from the intrinsic photoionization process.

Regarding dynamics, there is a fair fraction ($1-f_{\rm ellip}$) of BLR clouds in inflowing/outflowing motion.
The parameter $f_{\rm flow}$ is also not well constrained and have a large uncertainty for all models. 
Figure~\ref{fig_compare} demonstrates that for the H$\beta$ BLR in model M4, $f_{\rm flow}$ tends to be $<0.5$, 
which means that inflowing clouds are relatively preferred. This is consistent with the velocity-binned delay map obtained by 
\cite{Zhang2019}, which showed a longer time delay in blue velocity bins, a signature believed to indicate inflows
(\citealt{Peterson2014}).

\begin{figure*}[t!]
\centering 
\includegraphics[width=0.9\textwidth]{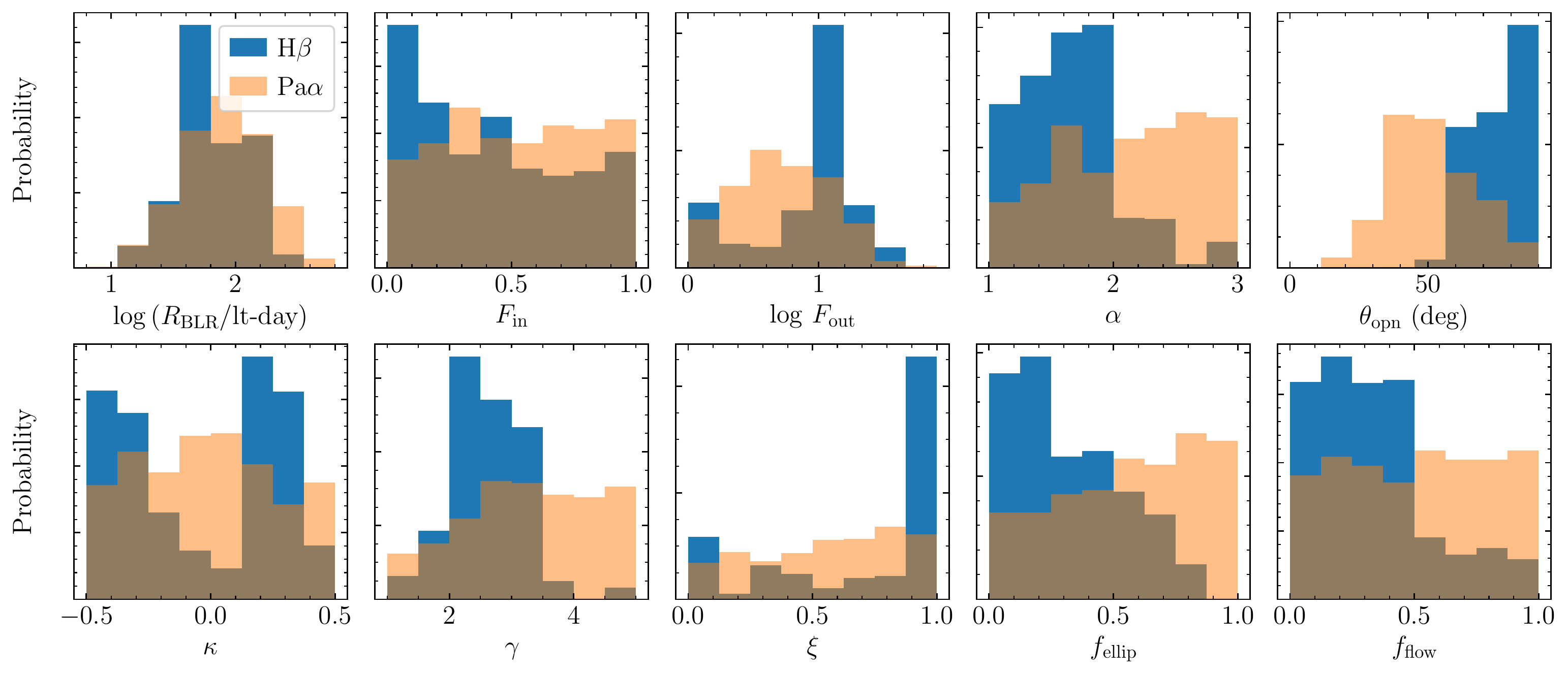}
\caption{The posterior distributions of major parameters of the most probable model M4 for the H$\beta$ and Pa$\alpha$ BLRs
(see Section~\ref{sec_bayes_infer} for the sampling covergence checking in diffusive nested sampling).}
\label{fig_compare}
\end{figure*}

\begin{figure*}
\centering 
\includegraphics[width=0.4\textwidth]{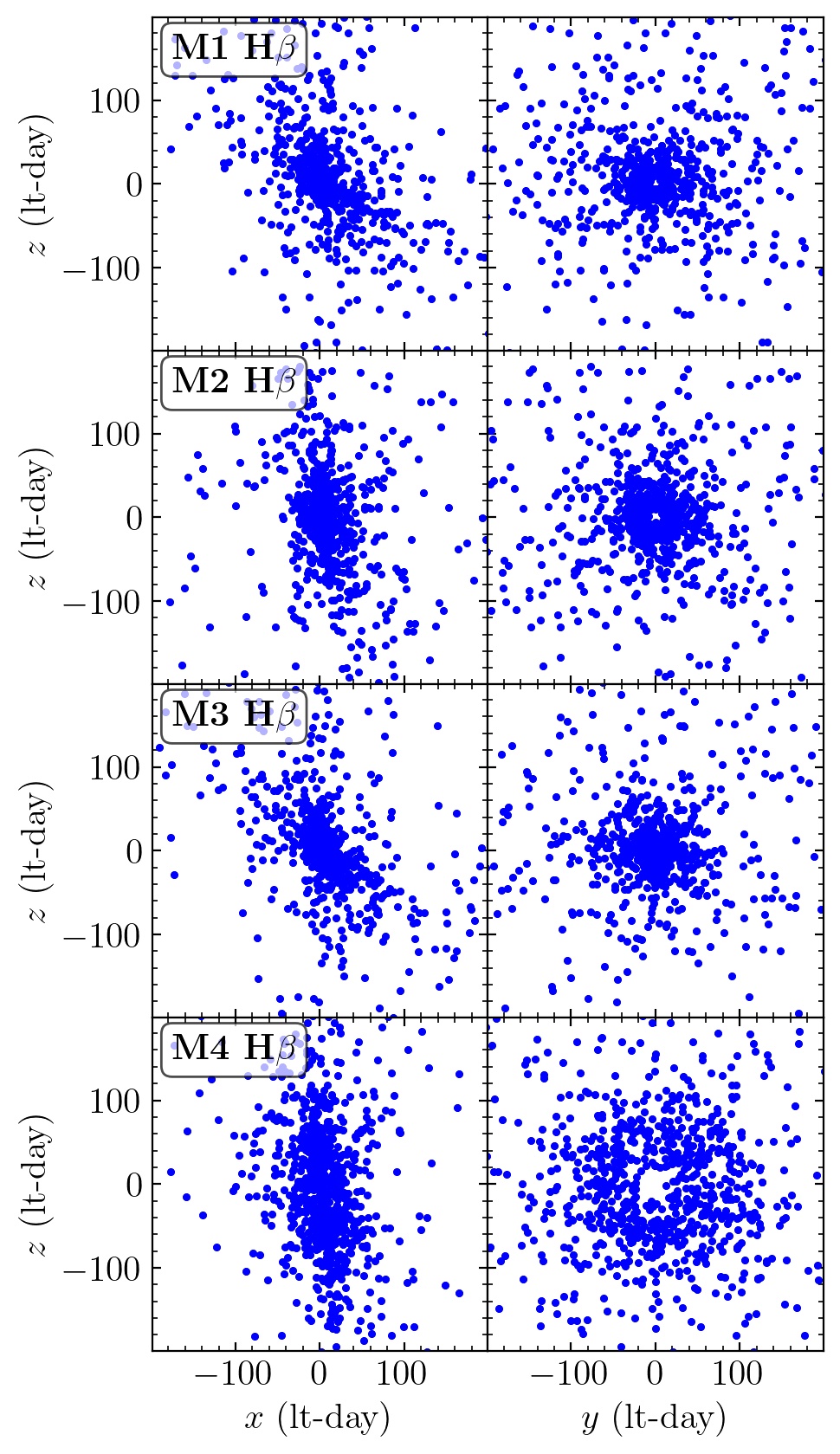}~~~
\includegraphics[width=0.4\textwidth]{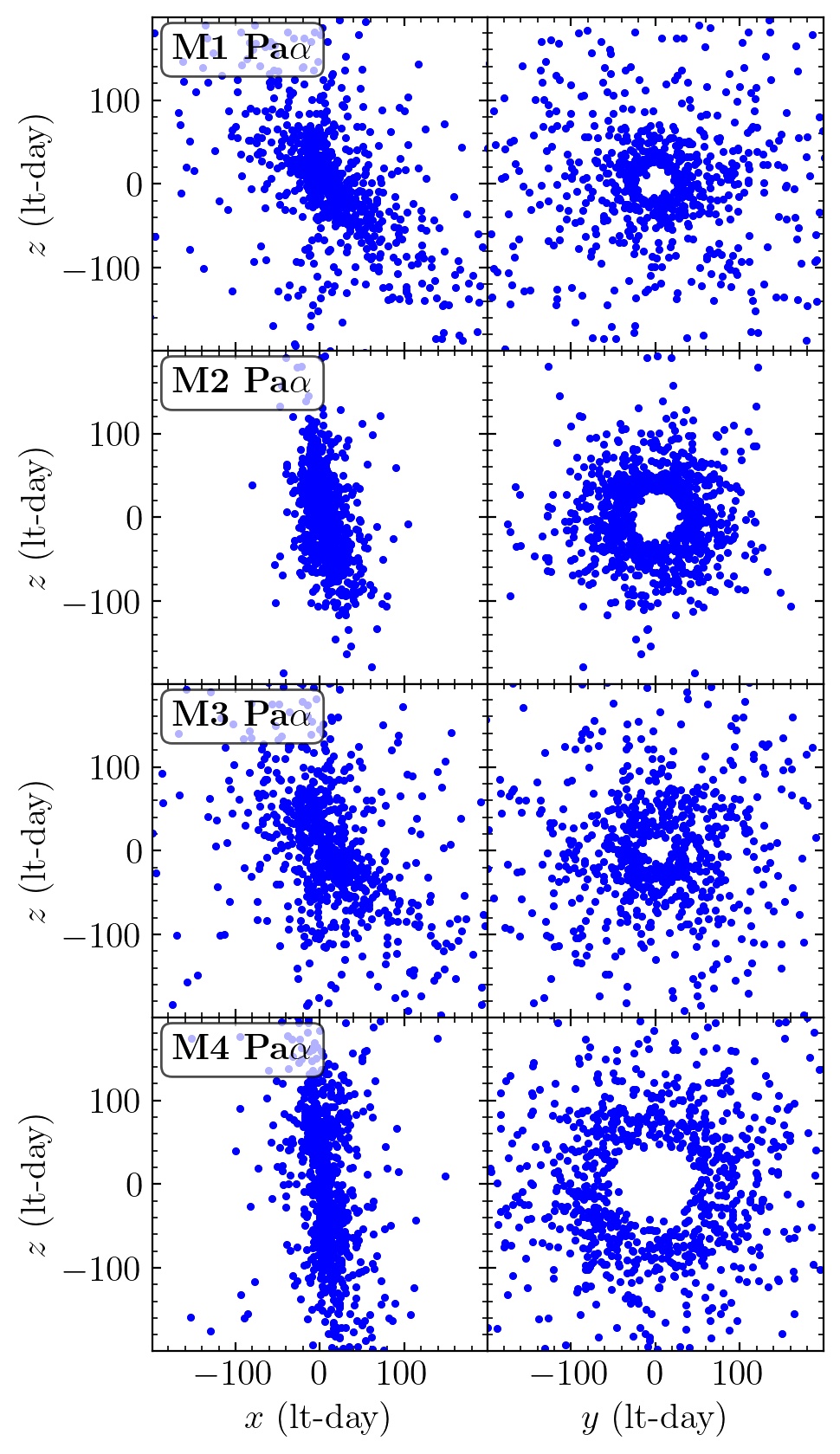}
\caption{Examples of cloud distributions of the H$\beta$ and Pa$\alpha$ BLRs for models M1-M4 
using the best parameter values listed in Table~\ref{tab_value}. 
The LOS is along the positive $x$-axis.}
\label{fig_clouds}
\end{figure*}

\section{Discussions}\label{sec_discussion}

\subsection{Compared to the SA Modeling of \citetalias{Gravity2018}}
\label{sec_comp1}

\citetalias{Gravity2018} applied a simple dynamical model to interpret the SA data of 3C 273.
In their model, BLR clouds follow a gamma distribution (D$r$1) in radial direction 
and a uniform distribution in vertical direction (D$\theta$2 with $\gamma=1$).
Such a geometry is somehow similar to that of model M2 and M6 if fixing $\gamma=1$.
However, for dynamics, \citetalias{Gravity2018} assumed circular Keplerian motion and did not include elliptical 
orbits and inflows/outflows. To determine the black hole mass from only SA data, 
\citetalias{Gravity2018} preassigned a value of $\log\,(D_{\rm A}/{\rm Mpc})=2.74$ for the angular-size distance of 3C~273,
roughly coincident with the inferred value of $2.61_{-0.30}^{+0.39}$ by model M2 and $2.71_{-0.35}^{+0.39}$ by model M6.
\citetalias{Gravity2018}\footnote{Note that the uncertainties in \citetalias{Gravity2018} were quoted 
with 90\% (1.65$\sigma$) confidence intervals. Here we simply divide the uncertainties by a factor of 1.65 to be consistent with
1$\sigma$ confidence intervals used in this work.} obtained a Pa$\alpha$ BLR size of 
$\log\,(R_{\rm BLR}/\text{lt-day})=2.16_{-0.06}^{+0.05}$ and a black hole mass 
of $\log\,(M_\bullet/10^8M_\odot)=0.41_{-0.13}^{+0.10}$. By a comparison, with model M2, our determined BLR size $\log\,(R_{\rm BLR}/\text{lt-day})=2.07_{-0.11}^{+0.22}$ and black hole mass $\log\,(M_\bullet/10^8M_\odot)=0.46_{-0.28}^{+0.51}$
(see Table~\ref{tab_value}). With model M6, the corresponding values are $\log\,(R_{\rm BLR}/\text{lt-day})=1.90_{-0.22}^{+0.34}$ and 
$\log\,(M_\bullet/10^8M_\odot)=0.46_{-0.57}^{+0.53}$. These quantities are consistent with the 
inferences of \citetalias{Gravity2018} within uncertainties.
Nevertheless, the much smaller uncertainties in \citetalias{Gravity2018} may be ascribed 
to the fixed angular-size distance $D_{\rm A}$ and a simple BLR dynamical model.

We note that our obtained inclination angle $\theta_{\rm inc}=8_{-3}^{+1}$ degrees by model M2 and $9_{-3}^{+6}$ degrees, slightly
smaller than the \citetalias{Gravity2018} value of $12\pm2$ degrees. This might be the main reason 
responsible for slightly larger black hole masses in our analysis.

\subsection{Compared to the 1D SARM Analysis of \citetalias{Wang2020}}
\label{sec_comp2}

As described above, \citetalias{Wang2020} conducted the first SARM analysis on the SA and 1D
RM data of 3C~273 following the same simple dynamical model in \citetalias{Gravity2018}. 
Because only using 1D RM data, \citetalias{Wang2020} assumed that 
the H$\beta$ and Pa$\alpha$ lines stem from exactly the same BLR and neglected 
the differences between the two lines by a zero-order approximation.
We relax this restrict of a common BLR for the H$\beta$ and Pa$\alpha$ by treating the two BLRs 
separately and let them only share the black hole mass and inclination.
To this end, 2D RM data are required. Such 2D SARM analysis is
more generally applicable, particularly in cases where SA and RM data are observed in different periods.

\citetalias{Wang2020} obtained a black hole mass of $\log\,(M_\bullet/10^8M_\odot)=0.75_{-0.07}^{+0.07}$,
angular-size distance of $\log\,(D_{\rm A}/{\rm Mpc})=2.74_{-0.07}^{+0.07}$, and inclination of 
$\theta_{\rm inc}=8.4_{-0.9}^{+1.0}$~degrees. The distance is consistent within uncertainties
with our obtained value by the most probable model M4 (see Table~\ref{tab_value}), however, the black hole mass 
is slightly lower than our inference. Again, this is because of a mildly larger inclination of \citetalias{Wang2020}
than ours (see Equation~\ref{eqn_mass_inc}).
We note that \citetalias{Wang2020} reported significantly smaller uncertainties of model parameters than ours.
The reasons are might be twofold. First, the assumption that the H$\beta$ and Pa$\alpha$ BLRs 
are identical leads to more strict constraints on model parameters. Second, compared to \citetalias{Wang2020},
we adopted a more generous BLR model that includes additional ingredients such as anisotropy of BLR emissions, 
transparency of the equatorial plane, elliptical orbits and inflow/outflows. These ingredients are necessary because 
the H$\beta$ profile are slightly asymmetric and the velocity-resolved delay map of the H$\beta$ line 
shows complicated structures (\citealt{Zhang2019}), which are unlikely to be reproduced with the simple BLR model
in \citetalias{Gravity2018}.

\subsection{Model Dependence}
Our results illustrate that the obtained BLR parameters (such as black hole mass and angular-size distance) 
are insensitive to specific geometric configurations and continuum detrending schemes, but 
more or less depend on parameterizations of the vertical distributions.
As mentioned in Section~\ref{sec_overview}, the vertical distribution D$\theta$1 (in which clouds tend to 
cluster close to the outer BLR faces) yields an inclination angle of $\theta_{\rm inc}\sim 30^\circ$
and black hole mass of $M_\bullet\sim10^8M_\odot$. By contrast, the distribution D$\theta$2  
systematically yields a smaller inclination angle of $\theta_{\rm inc}\sim 6^\circ$
and higher black hole mass of $M_\bullet\sim10^9M_\odot$.

As described in the preceding sections, the statistic tests with the Bayes factor and maximum 
likelihood also rank M4 as the most probable model.
The inclination angle and black hole mass obtained by 
M4 are all consistent with other independent constraints. 
This implies that it is possible to employ Bayesian statistics to select out the most probable model 
in cases without independent information. The Monte Carlo tests performed by \cite{Gravity2020a}
also confirmed the feasibility of using Bayesian statistics.
More cases of SARM analysis with independent observational constrains like 3C 273 presented 
in this work would testify the use of Bayesian model selection.

In addition, for AGNs with broad emission lines, polarized spectra are expected to
arise from the scattering electrons in the equatorial plane, which is effectively equivalent to 
an edge-on viewer (e.g., \citealt{Smith2002, Afanasiev2015, Songsheng2018}). It is thus possible that spectropolarimetry
of broad emission lines may provide independent information of the vertical distributions of BLRs, 
which is helpful to testify the results in the present SARM analysis. 
In this sense, spectropolarimetric observations merit investigations in future.

\subsection{Parameter Sharing of the H$\beta$ and Pa$\alpha$ BLRs}\label{sec_para_shareing}
In consideration of the different H$\beta$ and Pa$\alpha$ line shapes (see Figure~\ref{fig_prof}), 
we relax the restrict that the two lines have a common BLR and treat their respective BLRs 
separately. We let the two BLRs share the black hole mass and inclination angle.
This is the most conservative and unbiased implementation in so far as it is not yet clear
which factors are responsible for the noticeable differences in the two lines.
However, it is clear that any inappropriate parameter sharing of the two BLRs will 
cause biased results. The disadvantage of our present scheme of parameter sharing 
is that the obtained parameter uncertainties are relatively large, compared 
to full parameter sharing (see Section~\ref{sec_comp2}).
This disadvantage can be compensated by higher-quality RM data so that the black hole mass 
can be constrained more precisely.
Moreover, future better understanding of BLR geometry and kinematics by
including photoionization modeling may be beneficial to revealing the reasons for the 
different line shapes, therefore shedding light onto the parameter sharing.

\subsection{Joint Bayesian Analysis}\label{sec_joint}
In Table~\ref{tab_value}, we list $\chi^2$ of the best fits for the SA and RM data separately.
Because of much larger number of points in the RM data, the $\chi^2$ value is dominated by RM fitting,
which thus far out weights SA fitting in present recipe of joint analysis.
Meanwhile, the computational 
time for calculating H$\beta$ profile series is much intensive than that for calculating P$\alpha$ profile 
and differential phase curves. A possible procedure surmounting these issues is using the posterior from the SA fitting 
as the prior for the RM fitting (e.g., see \citealt{Alsing2021}). In generic, the posteriors might be 
complicated than single Gaussians. There is an increasingly popular technique, called normalizing flows,  that 
provides a general framework to transform any forms of posteriors to expressive probability distributions easy to deal with in 
MCMC sampling (see a review of \citealt{Papamakarios2021}). It is worth incorporating this technique to
our SARM analysis in future so as to expedite MCMC sampling.

\subsection{Possible Systematics and Future Improvements}
There are two remarks meriting further discussions in our calculations.

First, we assume that BLR clouds rotate coherently in a way that 
each cloud's angular momentum is diverse but overall
has a component oriented toward a common rotation axis.
SA signals sensitively depend on the degree of coherent motion of BLR clouds (e.g., \citealt{Songsheng2019}). 
This is because at a given velocity bin, the summation of the angular displacements will be canceled 
out once displacements are in mutually opposite directions (see Equation~\ref{eqn_tvt}).  In an extreme case, 
orbital motion of BLR clouds with fully random angular momentum would not produce SA signals.
The degree of coherent motion depends on specific formation scenarios of BLRs
(e.g., \citealt{Baskin2014,Wang2017, Czerny2017}). 
Other independent measurements about BLRs (such as spectropolarimetry) 
may be beneficial to justify the BLR scenarios and determine the degree of coherent motion.

Second, we include nonlinear responses of clouds' emissions to the incident continuum, described by
a parameter $\delta$. The conventional quantity, responsivity of clouds' emissions $\eta$, 
is related to the nonlinear parameter as $\eta = 1 + \delta$ (see Section~\ref{sec_assumptions}).
For simplicity, we assumed that $\delta$ and thereby $\eta$ are constant temporally and spatially so that 
the resulting emissivity-weighted and responsivity-weighted transfer functions 
are identical (see Section~\ref{sec_math}). 
In reality, photoionzation calculations shows that the responsivity depends on  
local gas properties (such as density and ionizing photon flux) and therefore may deviate from a constant 
(e.g., \citealt{Korista2004,Goad2014, Zhang2021}). A future improvement by incorporating photoionzation processes 
will  help to address this issue. On the other hand, the usual dispute that 
RM measures responsivity-weighted parts while SA measures emissivity-weighted parts of BLRs 
can be inherently resolved in present 2D SARM analysis. This is because we treat separately the BLRs 
for RM data and SA data. The connections between the two BLRs are the mutual SMBH mass and view inclination, 
rather than the mutual BLR size. Remarkably, such SARM analysis does not require SA and RM observations 
to undertake in the same periods and is also not affected by AGN intrinsic variability.
However, as mentioned in Section~\ref{sec_para_shareing}, this approach leads 
to a longer list of model parameters and thus larger uncertainties in parameter inferences.
On the theoretical side, better understandings of BLR dynamics and emission processes for different 
broad emission lines are required; on the observational side, high-fidelity data (namely,
with high signal to noise ratios and high and homogeneous cadences) are important to make improved 
measurements on black hole masses and distances.

\begin{figure}[ht!]
\centering 
\includegraphics[width=0.45\textwidth]{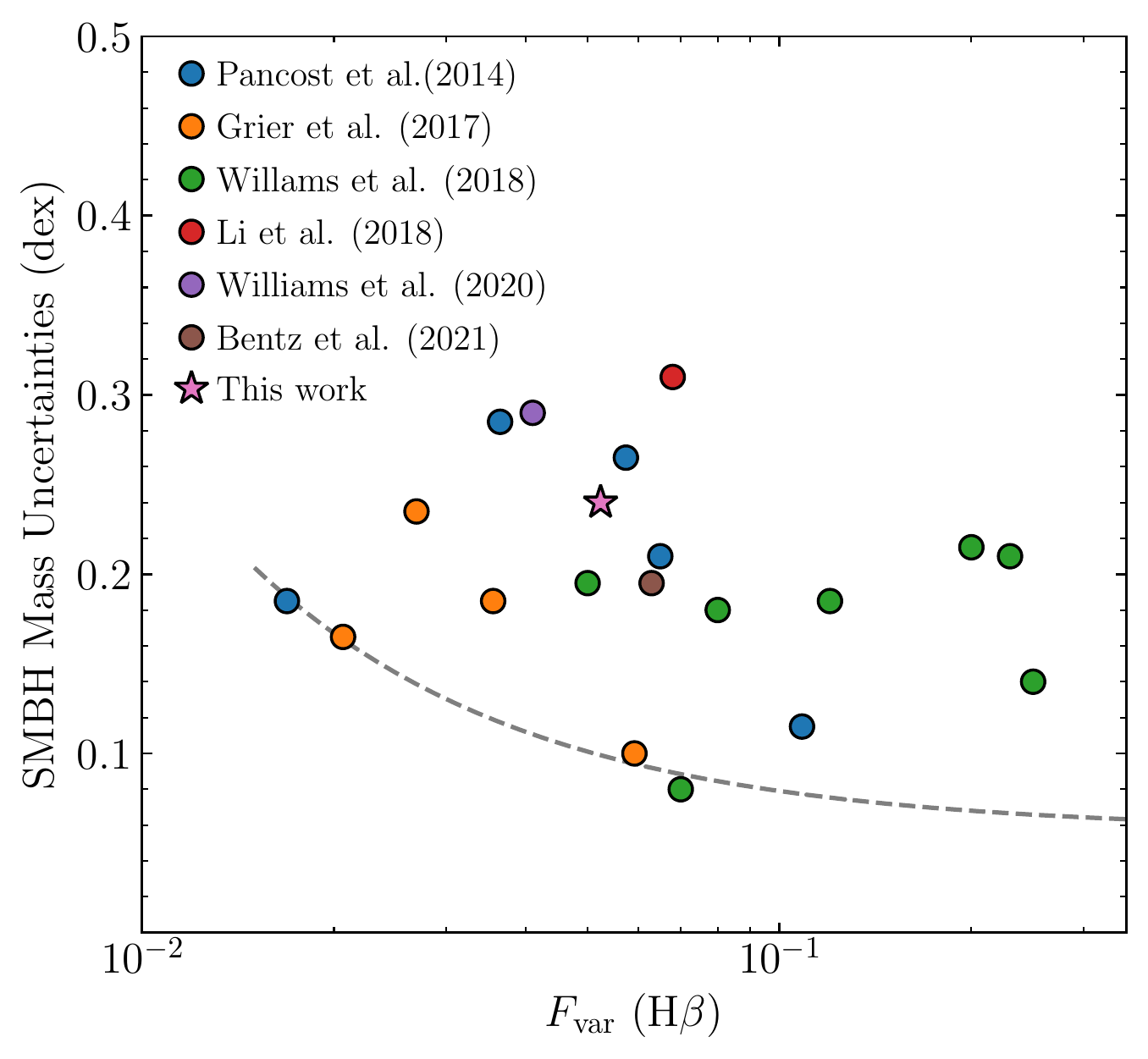}
\caption{The reported uncertainties of SMBH masses with fractional variability $F_{\rm var}$ of the H$\beta$ light curves 
from previous BLR dynamical modeling (see the text). The asterisk point represents 3C 273 in this work. The dashed line 
represents a simple extrapolation of minimum achievable uncertainties to large variability.}
\label{fig_err}
\end{figure}

\subsection{Measurement Uncertainties}
Our obtained SMBH mass has an uncertainty of about 0.24 dex, which translates to a large uncertainty of the angular-size 
distance (because the SA and RM data sets are linked through black hole mass and inclination). 
As mentioned in Section~\ref{sec_joint}, the $\chi^2$ value of
the joint analysis is dominated by the RM data. For the sake of comparison, in Figure~\ref{fig_err}, we compile the 
reported uncertainties of SMBH mass inferences with fractional variability of the H$\beta$ light curves from previous BLR dynamical modeling
with RM data (\citealt{Pancoast2014b, Gravity2017, Williams2018, Williams2020, Li2018}).
Here, fractional variability measures the excess variance of a light curve,  defined as $F_{\rm var}=\sqrt{\delta^2-\sigma^2}/\langle F\rangle$,
where $\delta^2$ is the variance of the fluxes, $\sigma^2$ is the mean square uncertainty, and $\langle F\rangle$ is the mean 
flux (\citealt{Rodrguez1997}). {\it If regardless of other influence factors} (such as cadences and time spans), 
a large $F_{\rm var}$ means that there are more significant variances in the light curves, which facilitates reverberation 
analysis and thereby leads to stronger constraints on parameter inferences. 
To guide eye, we use a line to extrapolate minimum achievable uncertainties to large variability in Figure~\ref{fig_err}.
With adequate variability and data quality, it seems possible to achieve a statistical uncertainty
of SMBH mass down to $<$0.1 dex (e.g., see \citealt{Williams2021}). 
For our present SARM analysis approach, a high-precision SMBH mass determination from the RM fitting effectively puts a strong prior to the 
SA fitting, therefore beneficial to yielding a competitive measurement on the angular-size distance.

\section{Conclusions}\label{sec_conclusion}
We construct a suite of BLR dynamical models with different geometric configurations
and conduct joint SARM analysis on the Pa$\alpha$ SA data and 2D H$\beta$ RM data  
of 3C~273. We consider two types of radial distributions and two types of vertical distributions.
In addition, a detrending procedure is needed to 
subtract the long-term trend in the continuum light curve (\citealt{Zhang2019}), which can be explained by the 
jet contaminations (\citealt{Li2020}). We use either a linear polynomial or the radio light curve of 3C~273
for the detrending procedure. We finally have 8 BLR models with different combinations of the above ingredients.
On account of of different  H$\beta$ and Pa$\alpha$ line profiles, 
we treat the their BLRs separately but let them only share the inclination angle and SMBH mass.
This allows us to determine the mass and angular-size distance of the SMBH in 3C~273 simultaneously in a self-consistent way.
Such a treatment has general applicability in cases where SA  and RM observations are undertaken in different periods.

We develop a Bayesian framework and employ a Markov-chain Monte Carlo
technique with the diffusive nested sampling algorithm (\citealt{Brewer2011})
to explore the posterior probabilities of the BLR dynamical models. 
Across the 8 models, the most probable model with
the maximum likelihood and Bayes factor is the one (M4 in Table~\ref{tab_models}) in which 
the radial distribution of BLR clouds is double power laws and the vertical distribution 
tends to concentrate around the equatorial plane. The obtained inclination angle 
is $5_{-1}^{+1}$ deg, coincident with the inclination of the large-scale 
jet in 3C~273 (see Section~\ref{sec_selection}). The best determined angular-size distance 
is $\log\,(D_{\rm A}/{\rm Mpc})=2.83_{-0.28}^{+0.32}$, and 
black hole mass is $\log\,(M_\bullet/M_\odot)=9.06_{-0.27}^{+0.21}$, 
which agrees with the relationships between black hole mass and bulge properties (see Figure~\ref{fig_mass_relation}).
Both the H$\beta$ and Pa$\alpha$ BLRs are geometrically thick and 
consist of a fraction of bound elliptical motion. The rest fraction are inflows or outflows, which cannot be distinguished 
with the present data, although inflows in the H$\beta$ BLRs are more preferable.

The final results are insensitive to the radial distribution of BLR clouds and 
the detrending approaches of the continuum light curve.
Nevertheless, we find that the obtained black hole mass and angular-size distance 
depend more or less on the vertical distribution of BLR clouds. Specifically, the distribution (D$\theta$1 
in Section~\ref{sec_geometry})
with BLR clouds clustering around the outer BLR faces results in a larger 
inclination angle and smaller black hole mass, just contrary to the distribution (D$\theta$2 
in Section~\ref{sec_geometry}) with BLR clouds concentrated towards the equatorial plane.
By comparing with other independent information about the inclination angle from 
the radio observations and the black hole mass from the relationships between black hole mass and bulge properties, 
our results implies that it is possible to use Bayesian statistics to determine the preferred model.

With the present data quality of 3C~273, the obtained black hole mass, angular-size distance, and other parameters
have relatively large uncertainties. Future high-fidelity SA and RM data (e.g., high signal to noise ratios and high 
and homogeneous cadences) may help to determine the BLR geometry and kinematics, black hole mass, and angular-size distance 
more preciously. Meanwhile, alternative independent measurements
on black hole mass such as through stellar/gas dynamics (e.g., \citealt{Kormendy2013, Barth2016, Bentz2016})
or through spectropolarimetry (e.g., \citealt{Afanasiev2015,Songsheng2018}) may 
also be valuable for this purpose.

%\acknowledgements{
\section*{Acknowledgments}
We thank the referee for useful suggestions that improve the manuscript and
the GRAVITY Collaboration for useful discussions and kindly sharing the GRAVITY observation data of 3C~273
through E. Sturm. We also thank Yulin Zhao for deriving the $K$-band bulge luminosity of 3C~273.
This work is supported by the National Key R\&D Program of China through grant No. 2016YFA0400701, 
by the National Science Foundation of China (NSFC) through grant No. 11833008, 11991051, and 11991054, 
by the China Manned Space Project though No. CMS-CSST-2021-A06 and CMS-CSST-2021-B11, and 
by the CAS International Partnership Program (113111KYSB20200014). Y.R.L. acknowledges financial support 
from NSFC through grant No. 11922304 and from the Youth Innovation Promotion Association CAS.
P.D. acknowledges financial support from NSFC through grant No. 12022301 and 11873048. C.H. acknowledges 
financial support from NSFC through grant No. 11773029. M.X. acknowledges financial support from NSFC through 
grant No. 12003036.

This research has made use of data from the Steward Observatory spectropolarimetric monitoring project, 
which is supported by Fermi Guest Investigator grants NNX08AW56G, NNX09AU10G, NNX12AO93G, and NNX15AU81G. 
This research has also made use of data from the OVRO 40 m monitoring program, which is supported in part 
by NASA grants NNX08AW31G, NNX11A043G, and NNX14AQ89G and NSF grants AST-0808050 and AST- 1109911.
%}

\software{\texttt{CDNest} (\citealt{Li2020cdnest}), \texttt{DASpec} (\url{https://github.com/PuDu-Astro/
DASpec}), \texttt{BRAINS} (\citealt{Li2018})}

\appendix
This appendix shows the full fits to the SA data of 3C 273 using models M4 and M8 in Figures~\ref{fig_fits_sa}
and \ref{fig_fits_sa_m8}.

\begin{figure*}[t!]
\centering 
\includegraphics[width=0.8\textwidth]{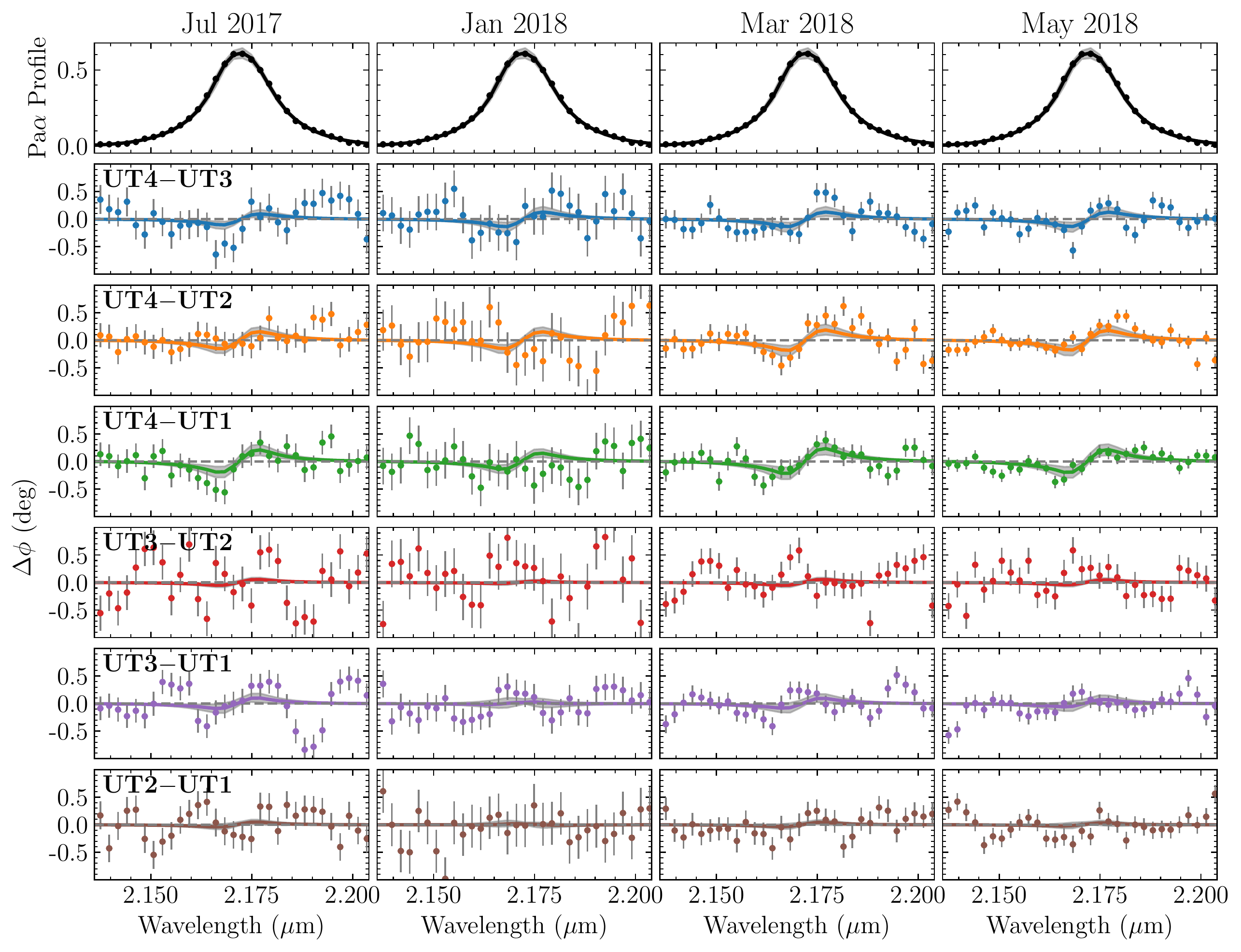}
\caption{Fits to the Pa$\alpha$ SA data of 3C 273 using model M4. The topmost four panels show the time-averaged Pa$\alpha$
profile (identical in each panel). The rest panels show the differential phase curves on the six 
baselines (rows) at four epochs (columns) observed by GRAVITY/VLTI (\citealt{Gravity2018}). 
The solid lines with grey shaded bands show reconstructions from model fits. The wavelengths are given in observed frame.}
\label{fig_fits_sa}
\end{figure*}

\begin{figure*}[th!]
\centering 
\includegraphics[width=0.8\textwidth]{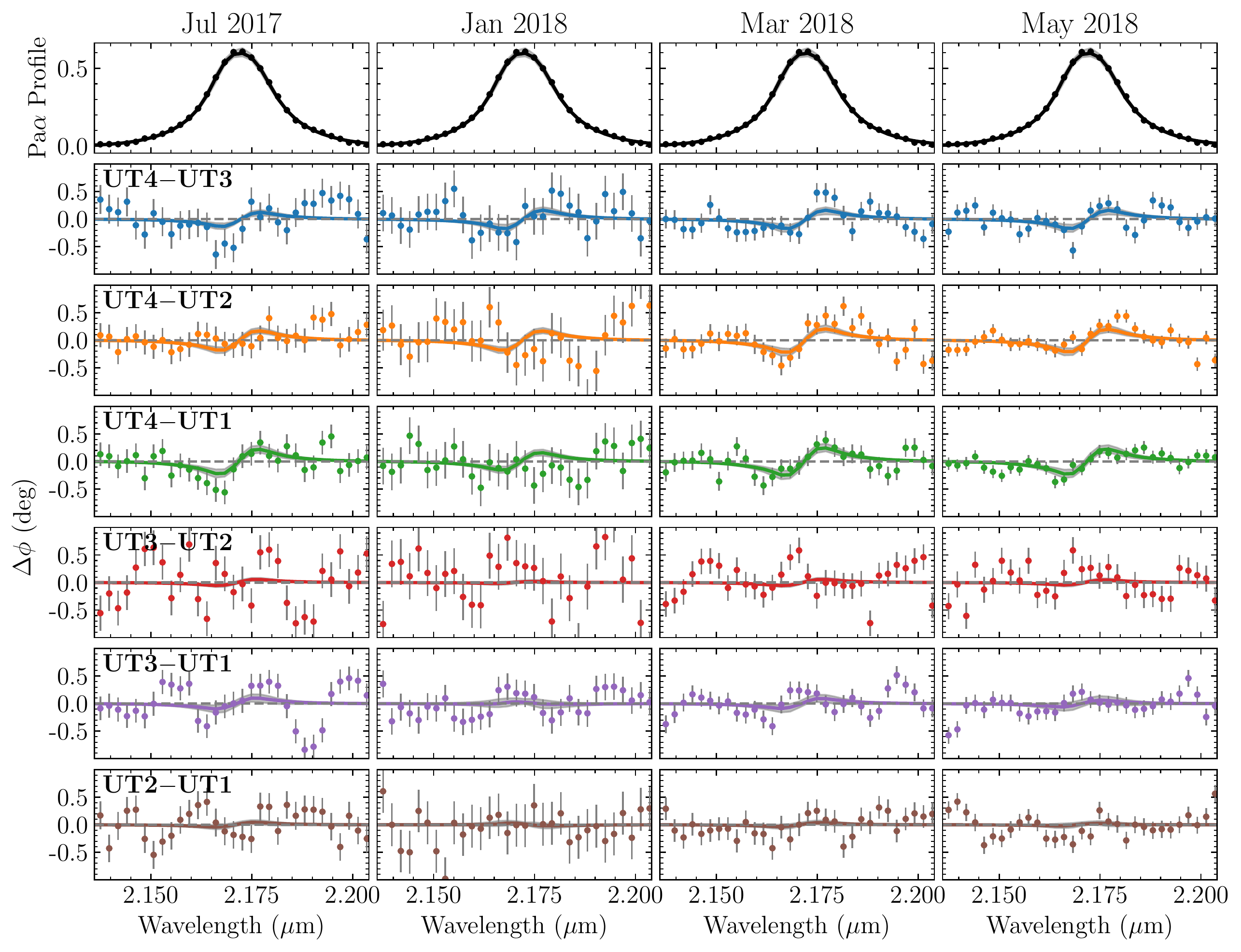}
\caption{Same as Figure~\ref{fig_fits_sa}, but using model M8.}
\label{fig_fits_sa_m8}
\end{figure*}

\end{document}